\documentclass[preprint]{aastex}
\usepackage{bm}

\usepackage{color}
\usepackage{graphicx}
\newcommand{\be}{\begin{equation}}
\newcommand{\ee}{\end{equation}}
\newcommand{\cov}{\mathrm{Cov}}
\begin{document}

\title{Measuring dark matter ellipticity of Abell 901/902 using Particle Based Lensing} 

\author{Sanghamitra Deb, David M. Goldberg} 
\affil{Department of Physics, Drexel University, Philadelphia PA-19104}
\author{Catherine Heymans}
\affil{University of Edinburgh, Royal Observatory, Blackford Hill, Edinburgh EH9 3HJ}
\author{Andrea Morandi}
\affil{Dark Cosmology Centre, Niels Bohr Institute, University of Copenhagen, Juliane Maries Vej 30, 2100 Copenhagen, Denmark} 

\email{sd365@drexel.edu}
\begin{abstract}
We present a non-parametric measure of the ellipticity and the alignment of the dark matter halos in Abell 901/902 supercluster. This super-cluster is a system of four separate peaks in a $0.5^{\circ}\times0.5^{\circ}$ field of view. We map the mass distribution of each individual peak using an improved version of Particle Based Lensing (PBL) and measure the ellipticity of the dark matter halos associated with two of the peaks directly from the mass map and by fitting them to a singular isothermal ellipse.  The parametric and non-parametric measurements are consistent for A901b while the position angle for the Southwest Group is different for the two techniques. We account for this discrepancy to substructure present in the Southwest Peak. We estimate an axis ratio of $0.37\pm 0.1$ for A901b and  $0.54^{+0.08}_{-0.09}$ for the Southwest Group.
\end{abstract}
\keywords{methods: statistical, methods:analytical, methods: data analysis, gravitational lensing,galaxies:clusters}

\maketitle

\section{Introduction}
Clusters of galaxies are the largest virialised structures in the universe.  Their mass distribution reveals properties of the primordial density field~\citep{1999MNRAS.308..119S,1990ApJ...351...10F,1974ApJ...187..425P}, and their mass function is sensitive to 
 predictions of  $\Lambda$CDM cosmology \citep{2009MNRAS.393L..31F,2006A&A...454...27B,1996MNRAS.282..263E}. As suggested by their low gas fraction, the baryons in clusters have little influence on the mass distribution beyond the core. Thus, clusters can be well approximated as dark matter halos. These halos produce spectacular arcs and multiple images  \citep{2009arXiv0907.4232Z, 2009arXiv0906.5079Z,2008A&A...489...23L, 2009arXiv0912.0916G} and  cause weak distortions of background galaxies.
  
Weak lensing inversion techniques study this distortion pattern to infer properties of the mass distribution of the cluster independent of its physical/dynamical state \citep{2006ApJ...648L.109C,  2008PASJ...60..345O}. Weak lensing has been used to measure dark matter density profiles in clusters \citep{2008ApJ...685L...9B,2005ApJ...619L.143B,2008ApJ...684..177U,2008JCAP...08..006M} to test predictions of N-body simulations of the standard $\Lambda$CDM model \citep{2007ApJS..170..377S, 2009ApJS..180..330K}. However, weak lensing has systematic and statistical errors. We have to understand and reduce these errors in order constrain the mass, size and shape of galaxy clusters.

Gravitational Lensing mass reconstruction of clusters have been studied for two decades \citep{2009arXiv0903.1103O, 2007MNRAS.374L..37K,2007ApJ...667...26P,2002ApJ...579..227I,2001AJ....121...10C,2000A&A...358...30V,1999ApJ...527..535W,1998MNRAS.296..392A,1995A&A...299..353W,1990ApJ...349L...1T,2009arXiv0901.4372O, 2007A&A...470..449B,2006ApJ...652..937B,2005ApJ...619L.143B,2001A&A...379..384C,2006MNRAS.365..414B} and many different methods have been developed in the process \citep{1995ApJ...439L...1K,1995A&A...297..287S,1995A&A...303..643B,1997MNRAS.286..696S,2001A&A...374..740S}. Almost all of them perform reasonably well in detecting massive peaks  (primarily associated with the Brightest Cluster Galaxy) in the field of view.  This is because massive peaks produce significant lensing signal which can be detected by the simplest weak lensing technique.The rapid improvement in the quality of observations has lead to extensive research on lens mass reconstruction techniques \citep{1998A&A...337..325S,1999MNRAS.302..118G, 2002MNRAS.335.1037M, 1999A&A...348...38L, 2007MNRAS.375..958D,2005A&A...437...39B,2009A&A...500..681M,2009ApJ...706.1201B, 2009arXiv0911.4972M}.

In this paper we have improved Particle Based Lensing (PBL) \citep{2008ApJ...687...39D}  by smoothing the ellipticities prior to mass reconstruction and evaluated the covariance of the resulting mass distribution. PBL is a mass reconstruction technique where the mass (or potential) is calculated at the location of each image. There is a weight function associated with each image with a kernel having a width that varies according to the local number density. 

We aim to measure the ellipticity of the dark matter halos of the 
super cluster Abell901/902. Abell 901/902 is a complicated system with several sub-clusters spread around a field of view of $0.5^\circ\times0.5^\circ(\sim 5\times5 \;\mathrm{Mpc^2})$ at a redshift of $z=0.165$.  This field was surveyed to study how galaxy evolution gets affected by the density of the environment \citep{2009MNRAS.393.1275G}. 
The mass and morphology of the dark matter halos of these galaxies play an important role in their response to surrounding environment. Since this field of view has three clusters and a group of galaxies it is a very good laboratory to study large scale structure and its influence on galaxy transformation. The weak lensing analysis ~(\cite{2008MNRAS.385.1431H}, hereafter H08) of this field reconstructs the large scale dark matter for this super cluster. The mass estimates of each peak were made using NFW profiles and non-parametric maximum likelihood methods. Comparison with the light profile has shown that the substructure in the dark matter peaks are closely followed by the substructure in galaxy groups. 

The paper is organized in the following way. In \S~\ref{sec:weak},~\ref{sec:opt},~\ref{sec:cov} we discuss the technical aspects of mass reconstruction and noise covariance analysis using PBL.  In \S~\ref{sec:ellip} we describe measuring ellipticity of the dark matter halo using parametric and non-parametric techniques. In \S~\ref{sec:data} we give a brief description of the data and in \S~\ref{sec:res} we report the estimated ellipticity for the individual peaks of the super cluster Abell901/902. In the last section we discuss our results and possible directions of future work.

\section{Weak Lensing Reconstructions}
\label{sec:weak}
Weak lensing is a statistical measure of small distortions in background galaxies caused by a cluster. 
The distribution of intrinsic ellipticities in source galaxies follows a Gaussian distribution with zero mean and a width of $0.2\sim0.3$~\citep{1999ARA&A..37..127M}. In order to extract a lensing signal from the weak distortion of image galaxies several of them have to be averaged to smooth out the intrinsic noise.
The ellipticities of the background image galaxies are the weak lensing observables. In the linear regime the ellipticity can be approximated by the shear of the lensing field. The shear has two components given by,
\begin{equation}
\gamma\equiv\gamma^{1}+i\gamma^{2}
\end{equation}
The shear $\gamma^{i}$ and the dimensionless surface mass density $\kappa$ are linear combinations of the second derivatives of the potential in angular coordinates,
\begin{eqnarray}
\label{eqn:gam1}
\gamma^{1}\equiv\frac{\psi_{,xx}-\psi_{,yy}}{2}\\
\label{eqn:gam2}
\gamma^{2}\equiv\psi_{,xy}\;\;\; \;\;\;\;\;\;\;\;  \\
\label{eqn:kap}
\kappa\equiv\frac{\psi_{,xx}+\psi_{,yy}}{2}\;\;\;
\end{eqnarray}
where $\psi_{,xx},\psi_{,xy}, ...$ refers to $\frac{\partial ^2\psi}{\partial^2 x},\frac{\partial ^2\psi}{\partial x \partial y}, ...$
In the weak lensing regime the ellipticity induced by the lens is given by,
\begin{equation}
\varepsilon^{i}=g^i=\frac{\gamma^{i}}{(1-\kappa)}
\label{eqn:ellipticity}
\end{equation}
and in the strong lensing regime the induced ellipticity is given by the inverse conjugate of the reduced shear. In this paper we will be addressing the weak lensing regime only. The above expressions are written for a source plane at infinity. The redshift dependence for convergence and shear are included
by,

\be
\kappa(\theta,z)=Z(z)\kappa(\theta) \;\;\;\;\;\; \gamma^{i}(\theta,z)=Z(z)\gamma^{i}(\theta)
\ee
where $Z(z)$ is the redshift weight function for background images given by,
\be
Z(z)=\frac{D_{ls}}{D_s}
\ee

\subsection{Non-parametric mass Reconstruction Techniques}
Non parametric weak lensing mass reconstruction techniques are broadly categorized into two categories.
There are direct techniques like KS93~\citep{1993ApJ...404..441K} and Finite Field Inversion~\citep{2001A&A...374..740S} where a linear relation is assumed between the measured ellipticity and the convergence. The advantage of these methods is that they are very fast and useful for testing the data. The disadvantage is that the linear techniques reconstruct the non-linear regimes less accurately. Hence the inverse techniques become important as we approach the strong lensing regime. 
In these methods a likelihood function is written between the observed ellipticties and the reduced shear and maximized iteratively to obtain the best solution.
\be
{\cal L}=exp\left[-\frac{\sum_{mn} \left({\varepsilon}^{i}_m-\frac{\gamma^i_m(\psi_m)}{1-\kappa_m(\psi_m)}\right)C^{-1}_{mn} \left({\varepsilon}^{i}_n-\frac{\gamma^i_n(\psi_n)}{1-\kappa_n(\psi_n)}\right)}{2}\right]
\label{eqn:chi0}
\ee
where $\varepsilon^{(i)}_{m,}$ are the observables. 
 Using finite differencing or Particle Based Lensing (PBL) the derivatives of the potential can be written as,
\begin{equation}
\psi^{(\nu)}_{n}=D^{(\nu)}_{nm}\psi_m
\label{eqn:D}
\end{equation}
where $\psi^{(\nu)}$ represents derivatives of the potential, $\nu$ is the order of the derivative. For first derivatives $\nu=x,y$, for second derivatives $\nu=(xx,xy \;\mathrm{and} \;yy)$ and so on. The likelihood function given by Equation~\ref{eqn:chi0} is linearized and written as a function of $D^{(\nu)}_{nm}$, the data correlation, the potential and the constraints at each step of the minimization and the potential is solved iteratively at the maxima of the likelihood.

\section{Method optimization}
\label{sec:opt}
Weak lensing is a statistical measure of small distortions in background galaxies caused by a cluster. 
 In order to extract a lensing signal from the weak distortion of image galaxies several of them have to be averaged to smooth out the intrinsic noise.  Smoothing is an integral part of any problem with noisy data and low signal-to-noise ratio. 
The error budget of weak lensing is controlled by the scale at which the data is smoothed.  Optimizing this scale is necessary independent of the technique that is used. In this section we will introduce smoothing, describe the inversion technique used to create mass maps from measured ellipticities and lay down a foundation for the choice of the optimal smoothing scale that will produce minimum reconstruction errors in the recovered mass.
 
In order to give equal weight to all ellipticity measurements around a single image we use an azimuthaly  symmetric smoothing function. We choose a normalized gaussian  for this purpose.
 \begin{equation}
\hat\varepsilon^{i}_{m}=Q_{mn}\varepsilon^{i}_n
\label{eqn:sm}
\end{equation}
where $\hat\varepsilon^{i}$ represents smoothed ellipticity field and $Q$ is the smoothing function. Here $m,n$ represent the background image index.
Smoothing of data prior to mass reconstruction has been done by several groups \citep{1995A&A...303..643B,1996A&A...305..383S,1998astro.ph..2051S,2001A&A...374..740S}, the covariance due to this smoothing is given by,
\begin{equation}
C_{mn}=Q_{km} Q_{kn} \sigma_n^2
\label{eqn:cov}
\end{equation}
Here $\sigma_n$ is the noise due to intrinsic ellipticities. 
This covariance between the constraints is used in the likelihood analysis in Equation~\ref{eqn:chi0}.

\subsection{Fitting the error}
\label{sec:fit}
We have defined a smoothing function and described the inversion procedure to create mass maps. The input parameter to this method is the smoothing scale. In this section we fit the weak lensing reconstruction error as a function of the smoothing scale, measurement error, number density and the length scale at which structure can be resolved.

\subsubsection{Toy Problem: A Sine Wave}
\label{sec:prob}

In order to understand how errors propagate into the reconstructed mass we will study a simple problem. The convergence $\kappa$ for this field is defined by,
\begin{equation}
\kappa=A \sin\left(\frac{2\pi x}{\lambda}\right)
\label{eqn:sin}
\end{equation}
where $\lambda$ is the wavelength, it represents the scale at which structure can be resolved. 
We do a weak lensing reconstruction of this wave and determine the scale at which the reconstruction error is minimal. We generate the mock data by adding noise to the shear (given by Equation~\ref{eqn:sin}) drawn from a Gaussian of width $\sigma_e$. This noise is varied over a wide range to determine a fitting formula for the error in the reconstructed $\kappa$.
We define the error that we are trying to measure as the difference between the model (Equation.~\ref{eqn:sin}) and the reconstructed mass at every image location (for particle based approaches like PBL)  averaged over the entire field of view. This error is given by,
\be
\sigma^2_{\kappa}=\langle(\kappa_{\mathrm{true}}-\kappa_{\mathrm{reconstructed}})^2\rangle
\label{eqn:def_err}
\ee

Here it is important to remember that while reconstructing $\kappa$ from data we will not be able to use Equation~\ref{eqn:def_err} as our goodness of fit since we do not know the underlying distribution of matter. In that case, the minimum $\chi^2$ will determine the best reconstructed mass.

We smooth the ellipticity field using a normalised gaussian of width $\zeta$,
The covariance introduced by smoothing data has been thoroughly studied by \cite{2003A&A...407..385L, 2002A&A...392.1153L, 2001A&A...373..359L}. We will construct an empirical form that fits Equation~(\ref{eqn:def_err}) by considering the effects of smoothing on mass reconstruction. The aim of smoothing is to average out the noise in the data, however, in this process we also wash out any structure smaller than the smoothing scale. 
 We take into account these two effects and fit them to the error in the reconstructed convergence from the mock data. 
\begin{equation}
\frac{\sigma_{\kappa}^2}{\langle\kappa^2\rangle}=A_0 \left({\zeta \over \lambda}\right)^4+B_0 \frac{\sigma_{\varepsilon}^2/(\langle\kappa^2\rangle n)}{(\zeta/\lambda)^{2} }+{C_0}
\label{eqn:err}
\end{equation}

Here $n$ is the number density per unit wavelength of background galaxy images. 
The first term represents the second order bias due to smoothing of small scale structure. The error introduced by this term is proportional to the square of the smoothing length.  The second term is the contribution from external error, in case of weak lensing this is the error due to the intrinsic ellipticities of the background galaxies. $ \sigma_{\varepsilon}^2/\langle\kappa^2\rangle $ represents the noise-to-signal ratio. The contribution from this term is inversely proportional to ${\left(\frac{\zeta}{\lambda}\right)^2n}$, implying that the external errors get washed out as the number density of images increases or as the area of smoothing increases. 

\begin{figure}[h]
\centering
\includegraphics[scale=0.55]{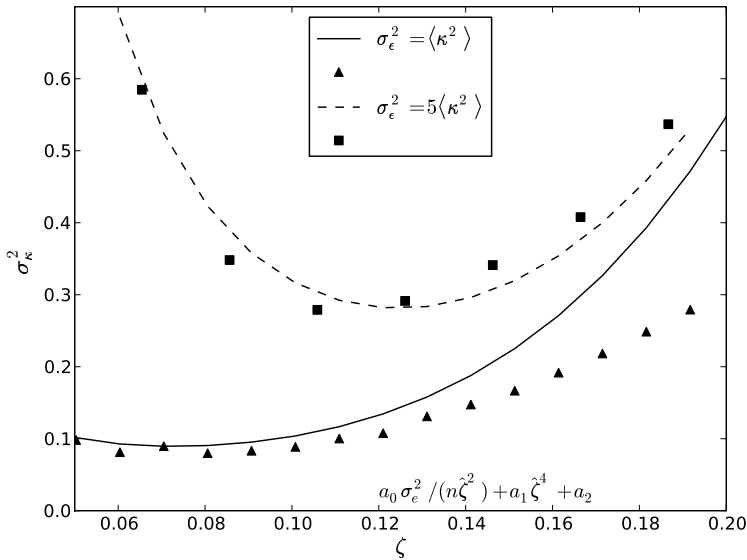}
\caption{Error vs the smoothing scale for two levels of noise. The triangles and the squares represent the errors measured from mass reconstruction done for different values of noise and smoothing scale. The solid line and the dashed line represent the fitted values for $\frac{\sigma^2_\epsilon}{\langle\kappa^2\rangle}=1,5$  given by Equation~\ref{eqn:err}, here $\hat\zeta={\zeta \over \lambda}$. The minima in this plot represents the ideal smoothing scale for the given noise. As expected decreasing the signal-to-noise for the mass reconstruction shifts the minimum towards higher smoothing scale.}
\label{fig:fit}
\end{figure}

Figure ~\ref{fig:fit} shows a fit for equation~\ref{eqn:err} to the observed error for increasing amount $\sigma_\varepsilon^2$.
 We have plotted two cases, the solid (dashed) line represents the fitted value and the triangles (squares) represent error for $\frac{\sigma^2_\varepsilon}{\langle\kappa^2\rangle}=1(5)$ measured from the reconstruction. Each data point (triangles and squares) on the graph represents a different mass reconstruction corresponding to a different smoothing scale and noise.
 As is evident from the plot Equation~\ref{eqn:err} is valid around the minima of the curves which defines the optimal smoothing scale for mass reconstruction.  This happens because Equation~\ref{eqn:err} was constructed by considering the first term in the Taylor expansion with non-zero contribution to errors. As we move away from the minima higher order terms start dominating. Since we are interested in computing the smoothing scale at which the error is minimum we do not need the higher order terms.
This is a demonstration with a toy model, we generalize this form and use Equation~\ref{eqn:err} to determine the scale at which we smooth ellipticities prior to the $\chi^2$ minimization.

\section{Covariance}
\label{sec:cov}
In this section we will compute the covariance due to smoothing in the reconstructed mass, which will be used to compute the ellipticity of the mass distribution. 
The covariance in the likelihood function has been studied extensively in the context of cosmic shear, \cite{2008arXiv0810.4254E} have studied the covariance between cosmological parameters to determine its effect on the likelihood analysis for cosmological parameter estimation. 
\cite{2008A&A...482....9E} uses the covariance among the data when comparing the information content in aperture mass and two-point correlation function. However, not much importance has been given to the role of the covariance in cluster cluster mass reconstructions with three exceptions of 
\cite{1998MNRAS.299..895B} who derive an analytic expression for the covariance and compare it with the covariances from Monte-Carlo simulations. Also \cite{2000MNRAS.313..524V} have shown that a maximum likelihood lensing mass reconstruction has a noise that follows a gaussian random field. \cite{2009A&A...500..681M} have also used covariance between the mass bins for computing mass maps. 

We have defined the covariance due to smoothing of external error in \S~\ref{sec:opt} in the ellipticities. 
The covariance of any linear function $f(\mathbf\theta)$ sampled at positions $\mathbf{\theta_n}$ between any two positions $\mathbf\theta_n$ and $\mathbf\theta_m$ is given by,
\begin{equation}
\cov(f;\theta_n,\theta_m)=\langle(f_n-\langle f \rangle)(f_m-\langle f \rangle)\rangle
\label{eqn:fcov}
\end{equation}
For the sake of simplicity of notation in the rest of the paper we will denote $
\cov(f;\theta_n,\theta_m) \equiv \cov^f_{nm}$, where $n,m$ represent either the
total number of image galaxies in case of PBL or the total number of grid cells in case of a grid based method.
We will derive the expression for the covariance matrix in the linear regime. This calculation is applicable to any technique where the derivatives of the potential can be expressed as a matrix times a potential. Hence it is applicable to PBL and finite-differencing techniques.
In the very weak lensing regime where $\kappa\ll 1$ the ellipticty can be approximated by the shear, $\gamma$,
\begin{equation}
\langle \varepsilon^{i}_n\rangle=\gamma^{i}_n
\end{equation}
where i=1, 2 represents the two components of ellipticity and shear. In this case we can write a likelihood between the observed ellipticty and the measured shear. Using Equation~\label{eqn:chi} we can write the likelihood as,

\begin{equation}
{\cal L}=exp\left[-\frac{\sum_{mn} \left({\hat\varepsilon}^{i}_m-G^{i}_{mp}\psi_p\right)C^{-1}_{mn} \left({\hat\varepsilon}^{i}_n-G^{i}_{nq}\psi_q\right)}{2}\right]
\end{equation}
 where $G^{i}$ is matrix that relates the potential to $\gamma^i$, C is the covariance due to smoothing and $ \hat{ \varepsilon}^{i}$ is the smoothed ellipticity field. Using Equation ~\ref{eqn:D}
 \begin{eqnarray}
 K&=&\frac{D^{xx}+D^{yy}}{2}\\
 G^1&=&\frac{D^{xx}-D^{yy}}{2}\\
 \;\;\;\;\;\;\;G^2&=&D^{xy}
 \end{eqnarray}
Setting the first derivative of the log of the likelihood to zero we get, 
\begin{equation}
\sum_{i}(G^{i})^{T}C^{-1} G^{i}\psi=\sum_i(G^{i})^{T}C^{-1} \hat{ \varepsilon}^{i}
\label{eqn:psi}
\end{equation}
From this equation we can solve for the potential, once we know the potential there is a linear relationship between the measured ellipticities and $\kappa$. In this case the estimated $\hat\kappa$ is given by,
\begin{equation}
\hat\kappa=K M^{-1}\sum_i(G^{i})^{T}C^{-1} \hat{ \varepsilon}^{i}=\sum_i V^{i} \hat{\varepsilon}^{i}
\label{eqn:kap_lin}
\end{equation}
where $K$ is matrix that relates the potential to $\kappa$, $M=\sum_{i}(G^{i})^{T}C^{-1} G^{i}$ and 
\begin{equation}
V^{i}=K M^{-1}(G^{i})^{T}C^{-1} 
\end{equation}
The covariance in $\hat\kappa$ follows Equation~\ref{eqn:fcov}. This can be re-written as,
\begin{equation}
\cov^\kappa=V\cov^{\hat{\varepsilon}}V^T
\label{eqn:cov_kap}
\end{equation}
where $\cov^{\hat{\varepsilon}}$ is the covariance in the ellipticity. There are several effects that contribute to this covariance. As discussed in \S~\ref{sec:fit} one term is due to smoothing of small scale structure and the other is due to averaging the error. We have derived the effect of smoothing on errors due to intrinsic alignment of galaxies  and used it for the maximum likelihood analysis. Here we are going to derive a more general expression for the covariance in the final reconstructed mass,
The covariance in measured ellipticity is given by,
\begin{equation} 
\cov^\varepsilon_{mn}=\langle(\varepsilon_m-\bar{\gamma}_m)(\varepsilon_n-\bar{\gamma}_n)\rangle=\delta_{mn}\sigma_n^2
\end{equation}
where $\bar{\gamma}_p$ is the true shear for the lens.
After smoothing this becomes, 
\begin{eqnarray}
\cov^{\hat\varepsilon}_{mn}&=&\langle(\hat{\varepsilon}_m-\bar{\gamma_m})(\hat{\varepsilon}_n-\bar{\gamma_n})\rangle\\\nonumber
&=&Q_{mp}Q_{np}(\sigma_p^2+\bar{\gamma_m }\bar{\gamma_n})+\bar{\gamma}_n\bar{\gamma}_m-\bar{\gamma_m }Q_{np}\bar{\gamma_p }-\bar{\gamma_n} Q_{mp}\bar{ \gamma_p }
\end{eqnarray}
Let us define,
\begin{equation}
\hat\gamma_m=Q_{mp}\bar{\gamma}_p
\end{equation}

Using this notation  we can write the above equations as,

\begin{equation}
\cov^{\hat\varepsilon}_{mn}=Q_{mp}Q_{np}\sigma_p^2+\langle(\bar{\gamma_m}-\hat{\gamma}_m)(\bar{\gamma_n}-\hat{\gamma}_n)\rangle
\end{equation}
The first term is same as Equation~\ref{eqn:cov}.
Since we do not know the true shear we replace $\bar{\gamma}$ with the reconstructed shear. Replacing this in Equation~\ref{eqn:cov_kap} we have an expression for the covariance in the linear regime. These relations are summarized in Table~\ref{tab:not}.

\begin{table}[h]
\begin{center}
  \begin{tabular}{| c | c | c | c | }
    \hline
  \hline
  $\cal{M}$& $\cal{X}$ & $\cal{Y=MX}$ & $\cov^{{\cal Y}} $
  \\ \hline
    \hline
   $ G^{1}$ & $\psi$ & $\gamma_1$ & $G^{(1)}\cov^{\psi}G^{(1)T}$\\ 
    \hline
     $ G^{2}$ & $\psi$ & $\gamma_2$ & $G^{(2)}\cov^{\psi}G^{(2)T}$
     \\ \hline
    $ K$ & $\psi$ & $\kappa$ & $K\cov^{\psi}K^{T}$\\ 
    \hline
     $ Q$ & $\gamma$ & $\hat\gamma$ & $-$\\ 
     $ Q$ & $\varepsilon$ & $\hat\varepsilon$ & $Q\sigma\sigma^{T}Q^{T}+\langle(\bar{\gamma_m}-\hat{\gamma}_m)(\bar{\gamma_n}-\hat{\gamma}_n)\rangle $\\ 
    \hline
     $ V$ & $\hat\varepsilon$ & $\kappa$ & $V\cov^{\hat{\varepsilon}}V^T$\\ 
    \hline
        \hline
       \end{tabular}
           \caption{ A summary of the relation between various matrices defined in\S~\ref{sec:cov}. The last column gives the covariance among the observable in the third column. The last row is the final expression for the covariance in the reconstructed $\kappa$.}
              \label{tab:not}
       \end{center}
       \end{table}

\section{Where is the information in Cluster Lensing? Cluster Ellipticity}
\label{sec:ellip}
It has long been established from simulations \citep{2002ApJ...574..538J,2006MNRAS.367..838R} that galaxy clusters are triaxial. 
\cite{2005ApJ...632..841O} have used triaxial dark matter halos to study the steep mass profile of A1689 and arc statistics using semi-analytic models~\citep{2003ApJ...599....7O} for triaxial dark matter halos have explained the abundance of gravitationally lensed arcs for a sample of clusters.
\cite{2004ApJ...613...95C} have measured the ellipticity and the position angle of a sample of X-ray selected clusters  parametrically using gravitational lensing  and found that the dark matter halos were aligned with the brightest cluster galaxies. In this paper we will compute the ellipticities of dark matter halos non-parametrically. This model-free estimation is done by calculating the quadrupole moments of the mass map using the noise matrix derived in the previous section. 

There is a strong dependence of ellipticity on amplitude of mass fluctuations $\sigma_8$ \citep{2006ApJ...647....8H}. A higher value of $\sigma_8$ indicates that clusters formation has started earlier and hence the measured ellipticity of clusters in the local universe would be lower. 
 Clusters are formed through hierarchical merging of smaller dark matter halos. Thus at their infancy they have more infalling matter and are more elliptical. As they virialise they become more and more spherical. Thus we expect clusters at higher redshift to be more elliptical than low redshift clusters. This has been confirmed by measuring higher order moments of the X-ray gas distribution  \citep{2005ApJ...624..606J,1995ApJ...452..522B}. 
  Following the procedure that we lay down in this paper we can measure the ellipticity of cluster halos from lensing for a large sample of clusters distributed in redshift. This will make the contribution of the errors due to projection smaller.  
 Since gravitational lensing probes the projected mass of the lens it is very difficult to constrain whether the halo is oblate or prolate.

We measure the shape of the lens by calculating the second order moments of the mass distribution, this will give us the eccentricity and the position angle for an elliptical mass distribution. We will also fit the ellipticity of the lens parametrically, for a truly elliptical lens this will give a very good description of the shape of the underlying dark matter halo. 

\subsection{Measuring Cluster Shapes Non-parametrically}
\label{sec:np}
In the previous sections we have derived the covariance of a mass map in the linear regime. The super cluster Abell 901/902 is a sub-critical cluster, hence for the error analysis we assume linearity. It is important to note that linearity is not assumed for doing the mass reconstruction. 
Since we have a correlated mass map with a covariance that is singular, we will perform a singular value decomposition as explained in  \S~\ref{sec:svd} and consider a few eigen modes that are significant. We then transform to a basis where the eigenmodes are independent. This is done by the transformation,

\be
\kappa^\prime=\langle\kappa\rangle+U^T(\kappa-\langle\kappa\rangle)
\ee

where U is the orthogonal matrix from singular value decomposition defined in Equation~\ref{eqn:svd}.
The $\kappa^\prime$'s are independent. We will express all quantities in terms of $\kappa^\prime$. Here
$\langle\kappa\rangle$ is the mean density of the field, it is calculated as follows

\be
\langle\kappa\rangle=\frac{\sum_{mn} C^{-1}_{mn} \kappa_m}{\sum_{mn}C^{-1}_{mn}}
\ee
The ellipticity of the dark matter is defined in terms of the quadrupole moments. The simple definition for  quadrupole moments for the $\kappa$ field is given by,

\begin{equation}
I_{ij}=\frac{\sum_{m}\kappa_m  w_m x^{(i)} x^{(j)}}{\sum_{m} \kappa_m }
\end{equation}

We know that $\kappa$ is correlated and hence we write the above expression in terms of $\kappa^\prime$ and inverse weight it by $s$, where $s$ is defined in Equation~\ref{eqn:svd}. The errors in the Quadrupole moments are given by the Octopole moments and are calculated in a similar fashion.
The ellipticities of the lens are defined by,

\begin{eqnarray}
e_1=\frac{(I_{xx}-I_{yy})}{I_{xx}+I_{yy}+(I_{xx}I_{yy}-I_{xy}^2)^{\frac{1}{2}}}
\end{eqnarray}

\begin{eqnarray}
e_2=\frac{I_{xy}}{I_{xx}+I_{yy}+(I_{xx}I_{yy}-I_{xy}^2)^{\frac{1}{2}}}
\end{eqnarray}

\subsection{Measuring shapes parametrically}
\label{sec:pm}
 We will fit singular isothermal ellipse to the dark matter peaks to constrain the shape of the dark matter halos. The convergence and the shear for this profile \citep{1994A&A...284..285K} is given by,
\be
\kappa=\gamma=\frac{1}{2}\frac{\theta_E}{\theta} f \left[\mathrm{cos}(\phi-\alpha)+f^2\mathrm{sin}(\phi-\alpha)\right]^{-\frac{1}{2}},
\label{eqn:par_e}
\ee

where $\theta_E $ is the Einstein radius given by,

\be
\theta_E=4 \pi \frac{\sigma_v^2}{c^2}\frac{D_{ls}}{D_s},
\ee
and $\sigma_v$ is the velocity dispersion of the cluster.
 $f$ is the axis ratio and $\alpha$ is the position angle. This relationship is degenerate with $\alpha=\alpha+\pi/2$ and $f=1/f$. The complex ellipticty is related to the to axis ratio via,

\be
e=\left(\frac{1-f}{1+f}\right) e^{2i\alpha}
\ee

This model is fitted by minimizing a $\chi^2$ of the form

\be
\chi^2=\sum_{m,i}\frac{(\varepsilon^{i}_{m}-g^{i}_{m})^2}{\sigma_{\varepsilon}^2}
\ee
where $i=1,2$ the two components of ellipticity, $\varepsilon^{i}_{m}$ is the ellipticity of the mth galaxy and $g^{i}_{m}$ is the reduced shear  and  $\sigma_{\varepsilon}$ is the error in the tangential ellipticity.
 This form of the mass model has been investigated by 
\cite{2001A&A...369....1K} and applied to a sample of X-ray luminous clusters \citep{2004ApJ...613...95C}.

\section{Data}
\label{sec:data}
The shear catalog for this cluster is generated following algorithms described in \cite{1995ApJ...449..460K, 2006MNRAS.371L..60H}. The modeling of the temporal variation of the PSF is outlined in \cite{2005MNRAS.361..160H}. The Abell901/902 field of view has several tens of good stars for stellar modelling of the PSF, the stars were chosen to maximize the signal-to-noise of the stellar ellipticity function and minimize the temporal variation. The charge transfer effieciency of the ACS has degraded over the years, this is corrected using the methods proposed by \cite{2007ApJS..172..203R}. Furthermore magnitude cuts are applied to the galaxy catalog to eliminate cluster members and foreground galaxies. The magnitudes are chosen such that $23<m_{F606W}<27.5$ and the signal to noise for each galaxy is chosen to be $>5$ with a size greater than 3 pixels. The total sample size is $\sim60000$ galaxies with 65 galaxies per square arcminute. For Abell901/902 the redshift for 90\% of the source galaxies were not known. However, the redshift weight function does not have a very strong dependence on the redshift for $z_s>1$ and $z_l=0.165$.
 Hence we assume the sources to be at a single redshift of 1.4 which is the estimated median redshift of the background sources. A more detailed description of the data and tests for systematics can be found in H08.

\section{Results}
\label{sec:res}
\begin{figure}[h!]
\centering
\includegraphics[scale=0.5]{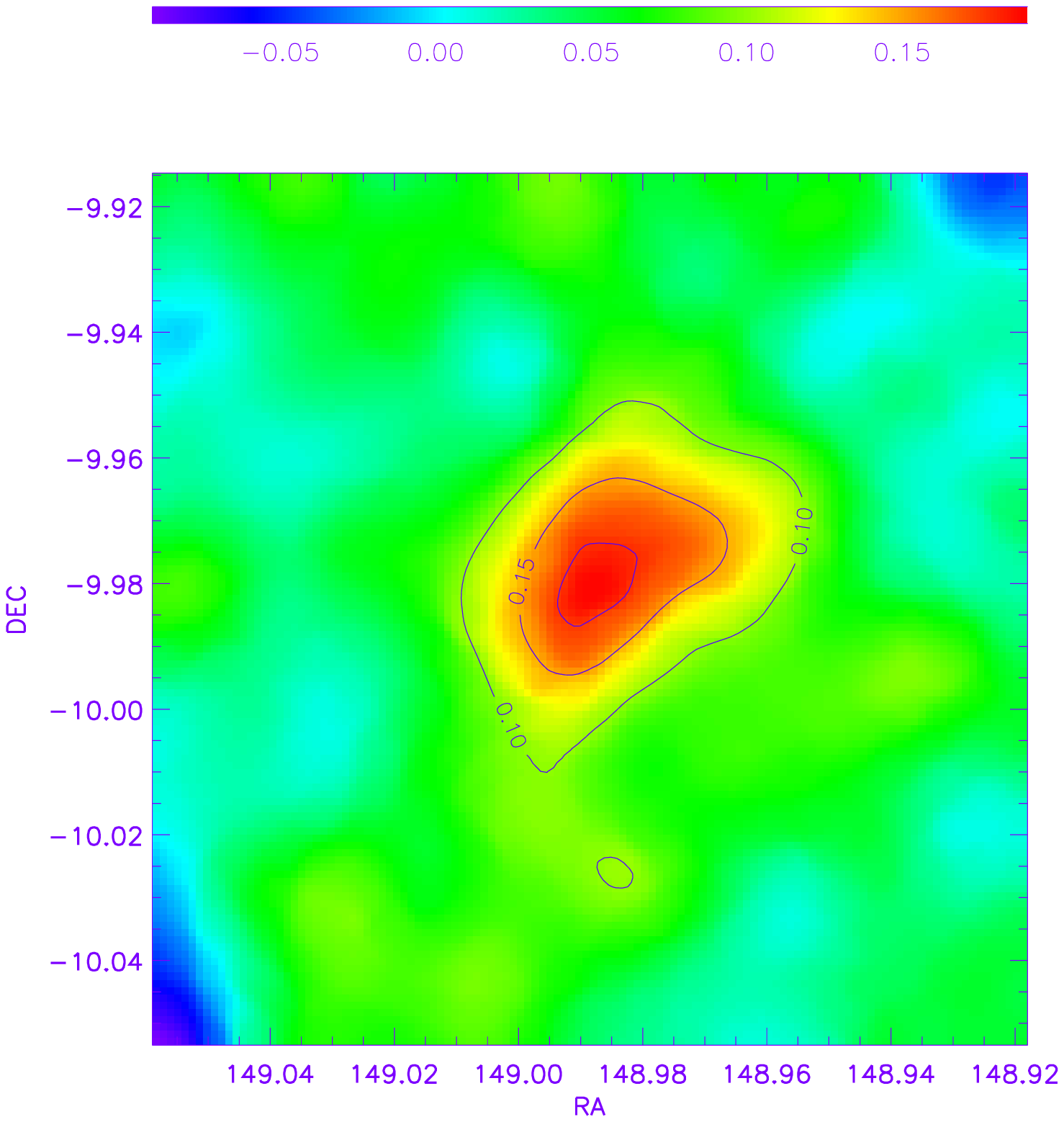}\includegraphics[scale=0.5]{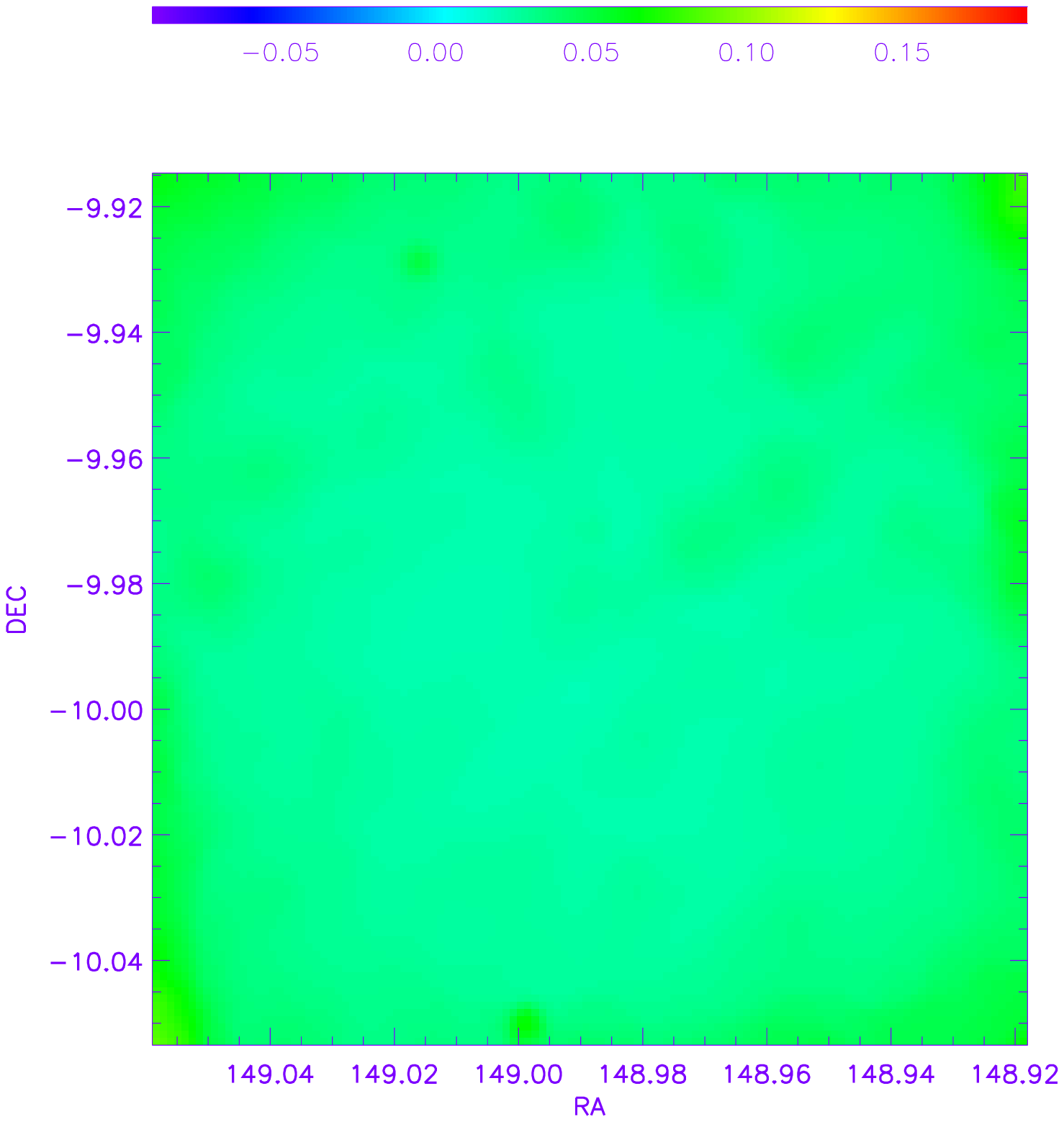}
\includegraphics[scale=0.5]{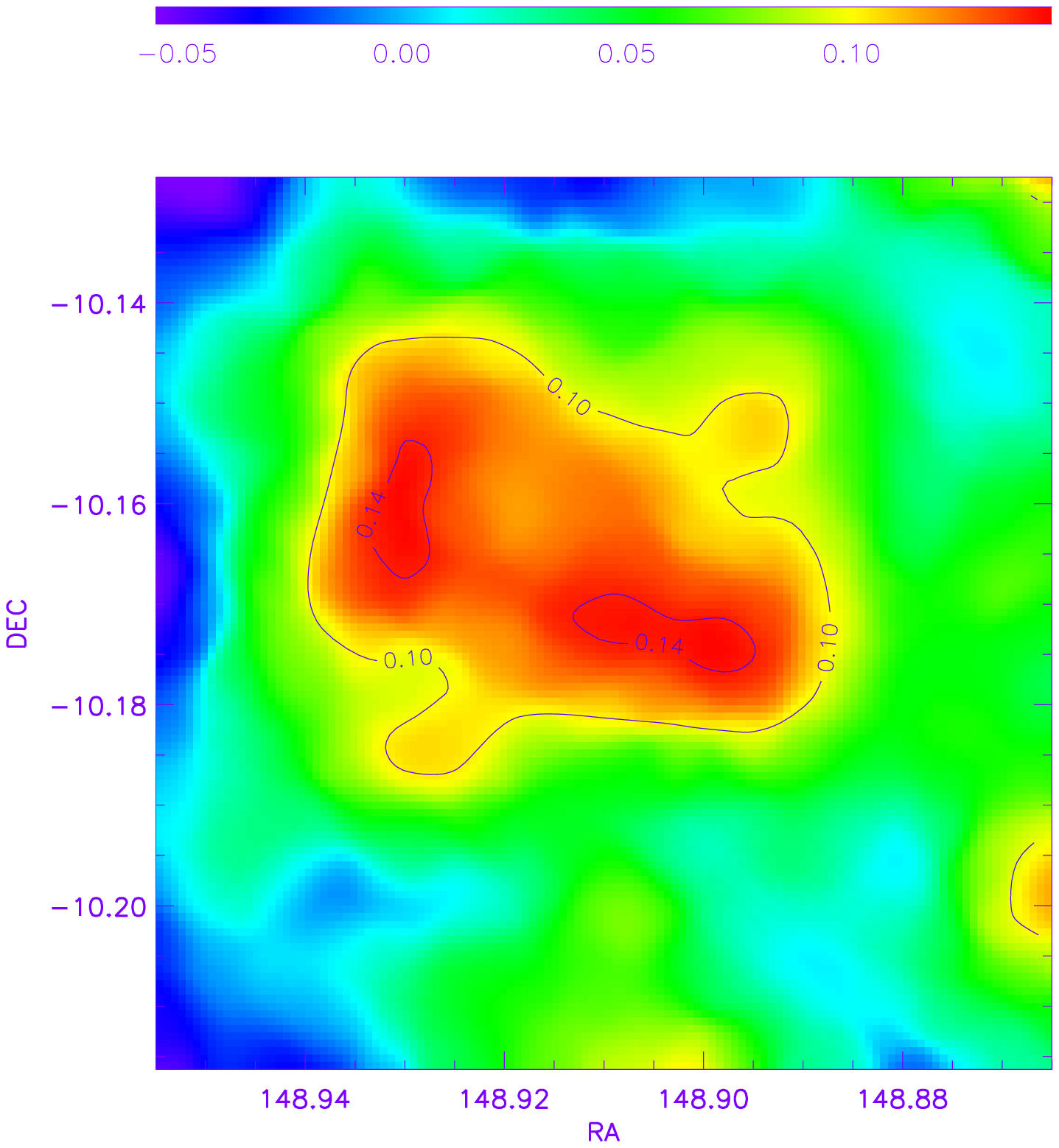}\includegraphics[scale=0.5]{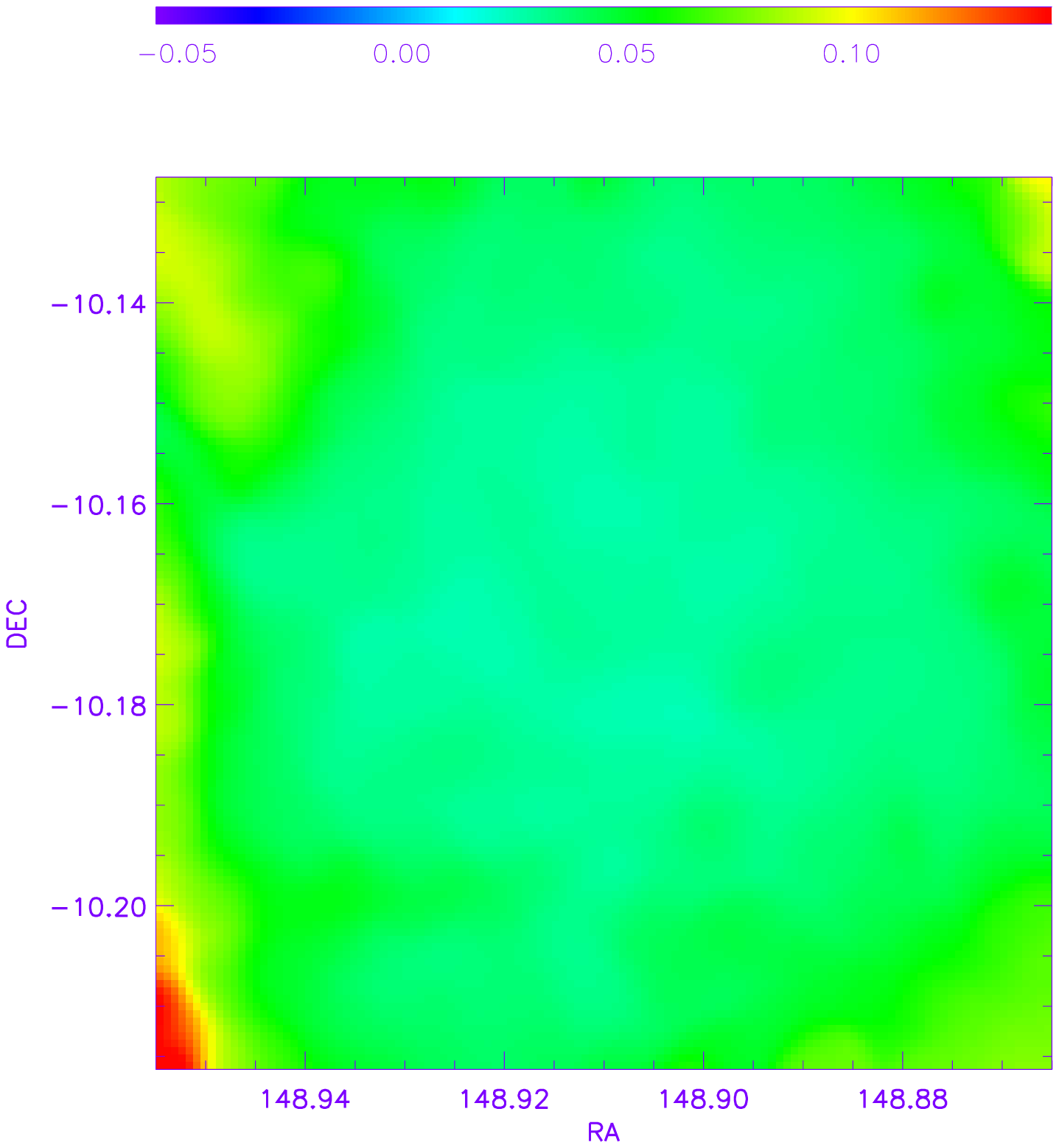}
\caption{ Mass reconstruction of dark matter halos in A901/902. Top Left Panel: A901b Top Right Panel: Error Map of A901b. Bottom Left Panel: SouthWest Group. Bottom Right Panel: Error Map for the Southewest Group. There is an artificial increase of error towards the extemities  of the maps where the convolution kernel steps over its hard edges.A901b is a compact dark matter halo, the peak is detected at $7\sigma$  and the Southwest Group has significant substructure with two sub-peaks detected at $4\sigma$ . We measure the ellipticity of these two peaks.}
\label{fig:sw}
\end{figure}

We have outlined a recipe for choosing the smoothing scale and using PBL for mass reconstruction. We will
apply this technique to the Abell 901/902 field of view. The dark matter peaks in this field of view has been detected by \cite{2008MNRAS.385.1431H}. We will zoom into each of the peaks and reconstruct a mass map and an error map for the peaks. For these peaks the signal-to-noise ratio $\sim 0.3$, using this value in Equation~\ref{eqn:err} we get an approximate smoothing scale of $0.5^\prime$.
We reconstruct the dark matter distribution of the four peaks. 
First we will study A901b, it is the most compact peak with a single halo. The minor to major axis ratio and position angle is reported in Table~\ref{tb:data_ellip}. The ellipticity of the peak is $0.45^{+0.11}_{
-0.10}$ and the position angle is $90^{\circ}$ suggesting that it is elongated vertically. The top left panel of Figure~\ref{fig:sw} is a map of A901b. 

\begin{figure}[t!]
\centering
\includegraphics[scale=0.5]{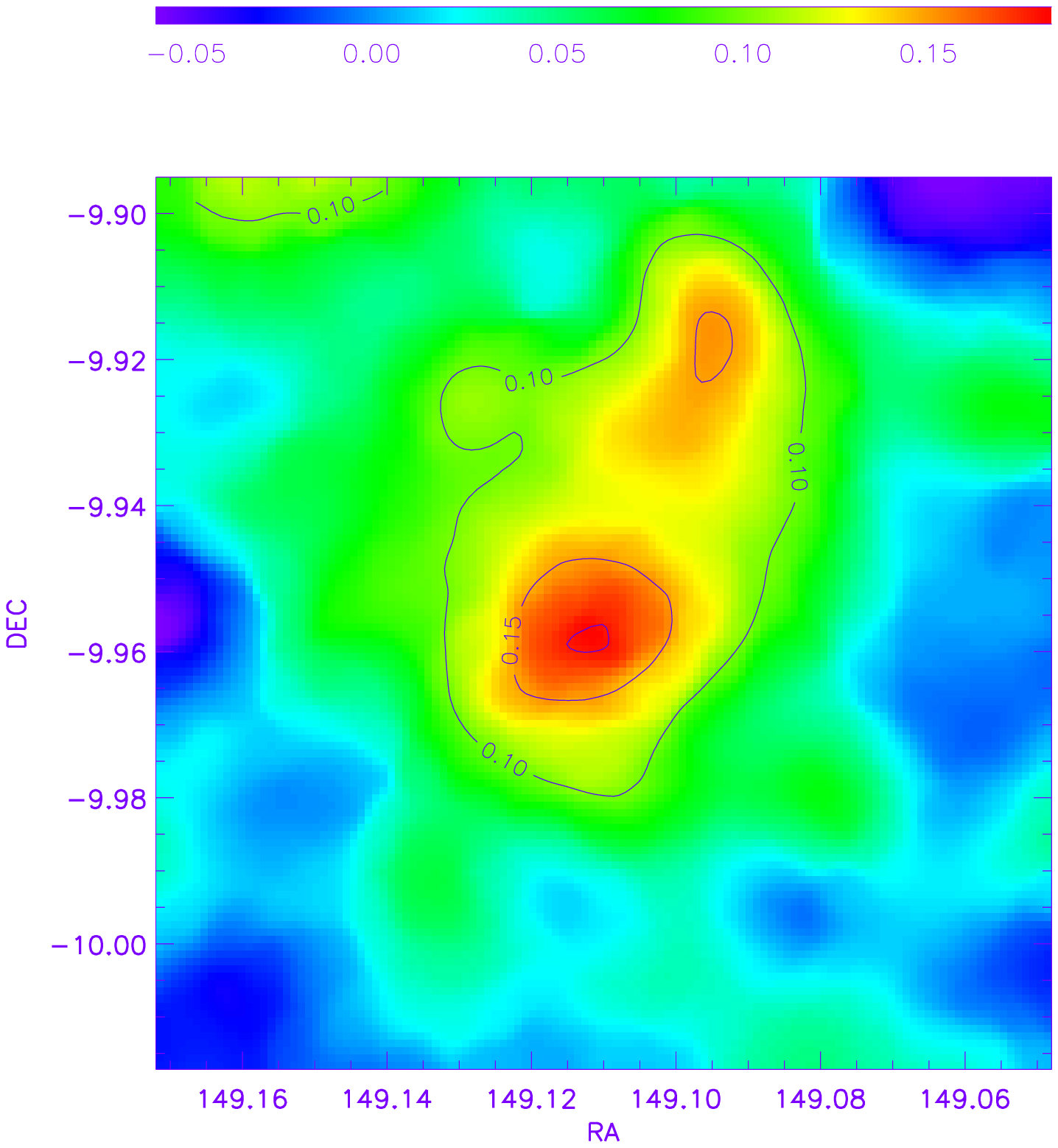}\includegraphics[scale=0.5]{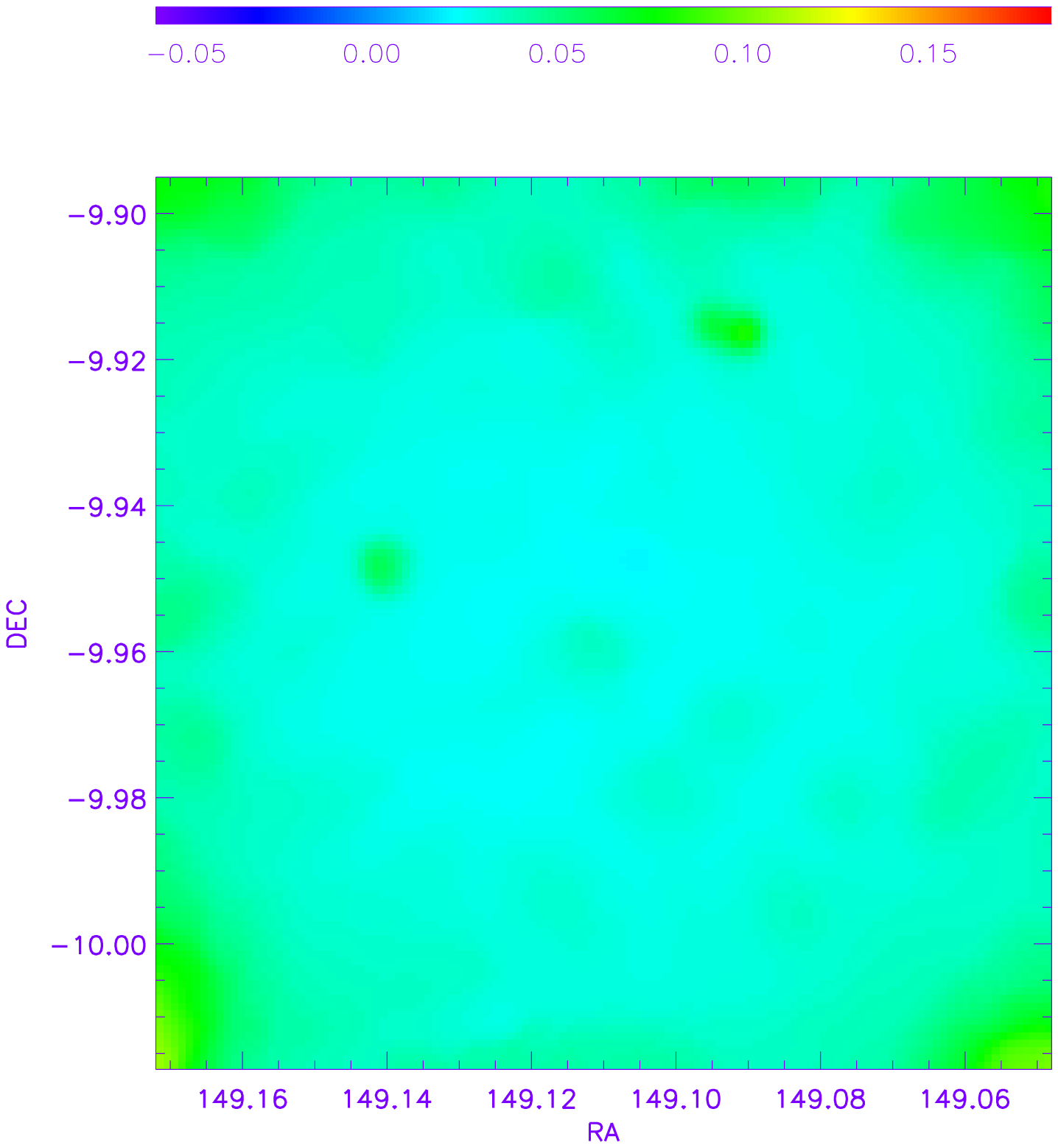}
\includegraphics[scale=0.5]{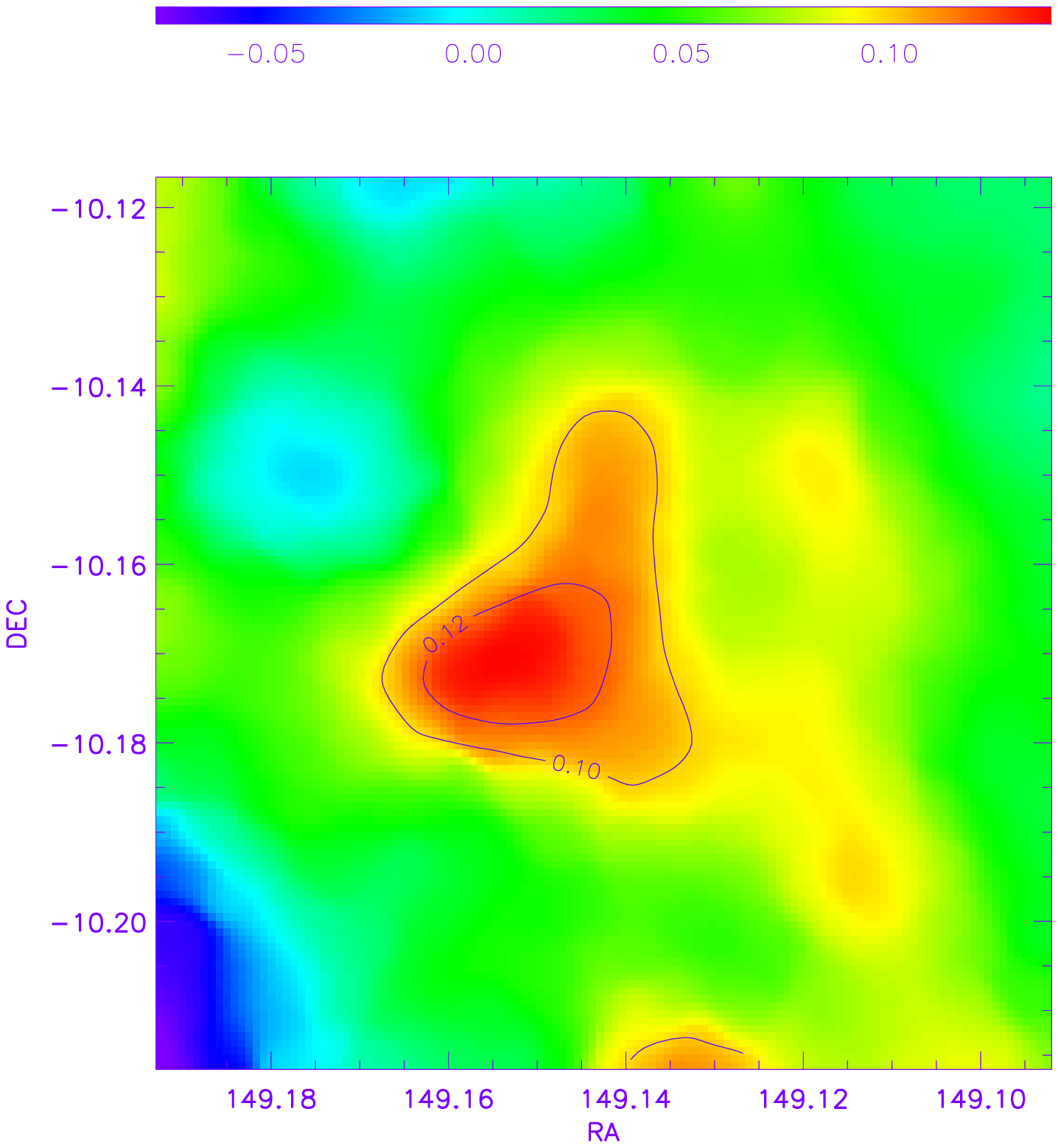}\includegraphics[scale=0.5]{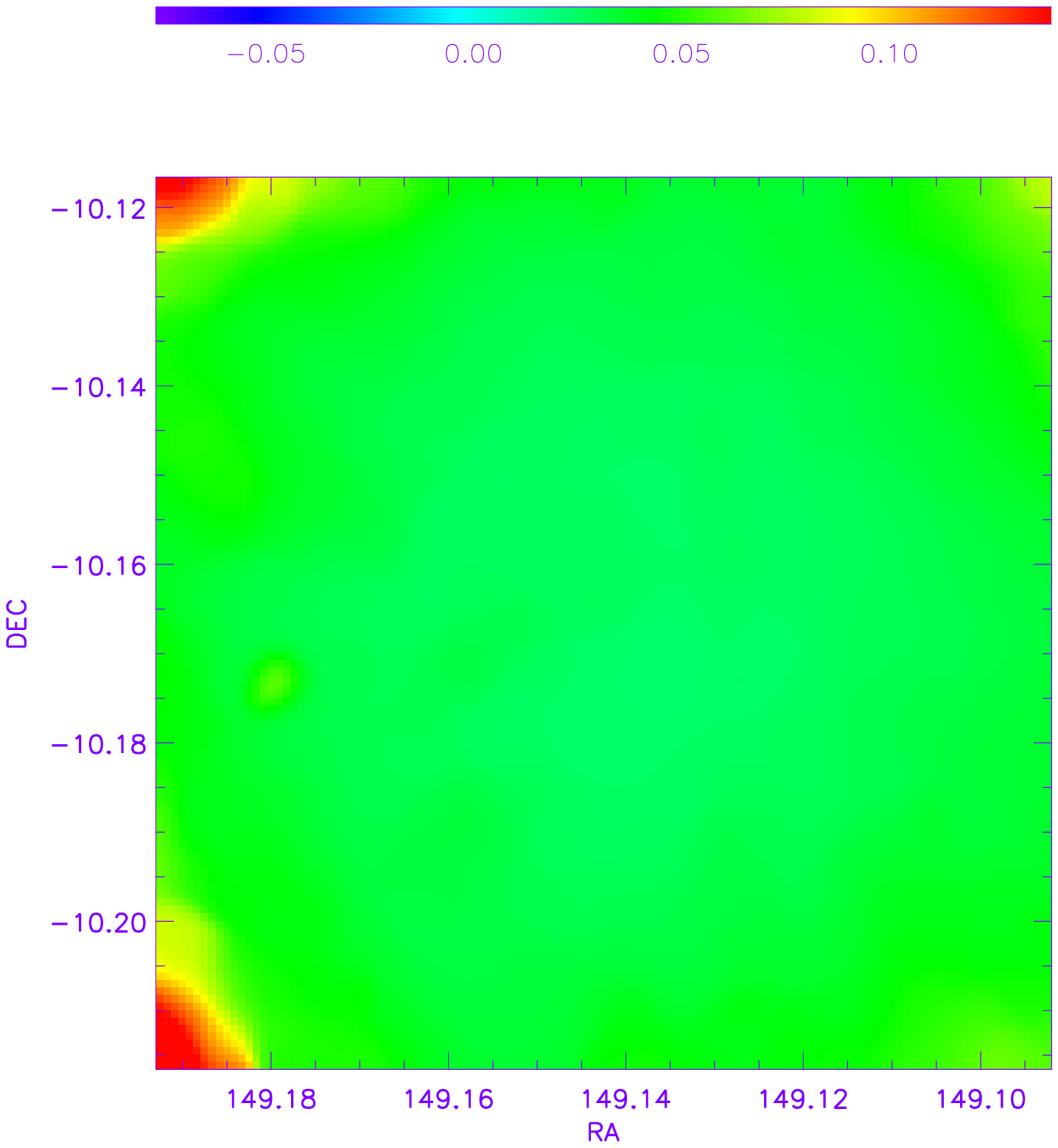}
\caption{Mass reconstruction of dark matter halos in A901/902. Top Left Panel: A901a Top Right Panel: Error Map of A901a. Bottom Left Panel: A902. Bottom Right Panel: Error Map A902.  A901a has two distinct peak, the central peak is detected at $5\sigma$ and the secondary peak is detected at $2\sigma$.  A902 is reasonably disturbed with the central peak detected at $4\sigma$. These two peaks are not representative of elliptical dark matter halos.}
\label{fig:a901}
\end{figure}

The bottom left panel of Figure~\ref{fig:sw} is the dark matter reconstruction of the Southwest Group.
This peak clearly shows that it is not spherically symmetric, it is elongated in the north west direction with significant substructure suggesting that the group is not completely virialized.  We measure an axis ratio of $0.3\pm0.07$  and a position angle of $120^\circ\pm 4.8^{\circ}$. Since the dark matter distribution has multiple maxima an elliptical dark matter halo may not be the best description for the Southwest Group. 
 A901a has two distinct peaks. This is clear from the dark matter map of the top left panel of Figure~\ref{fig:a901}. It has non-zero quadrupole moments, however, it is not possible to describe it using an ellipse.
 
 If we compare Figure~\ref{fig:sw},\ref{fig:a901} with H08 we see that the dark matter distribution is very similar. However, in this analysis we have smoothed the data prior to reconstruction and included the covariance due to smoothing in the $\chi^2$ minimization and in calculation of covariance of the final mass reconstruction. This makes the errors of the reconstruction well understood. The error maps in the right hand panels of Figure~\ref{fig:sw}, \ref{fig:a901} is computed by taking the square root of the diagonal of the covariance matrix of $\kappa$. The central peak of A901b is detected at $7\sigma$ significance, the two sub-peaks of the Southwest Group is detected at $4\sigma$. The central peak of A901a is a $5\sigma$ detection the secondary peak is a $2\sigma$ detection. The peak of A902 is detected at $4\sigma$.
The mass measured within one arcminute of the center of each peak is listed in Table~\ref{tb:data_ellip}. We measure the ellipticity of A901b and the Southwest peak since these two peaks are relatively less disturbed. A901a has two distinct peaks and A902 is very disturbed even in the outer regions. Hence it is difficult to represent these peaks with ellipses.
 
\subsection{Parametric Fitting}
\begin{figure}[h!]
\centering
\includegraphics[scale=0.3]{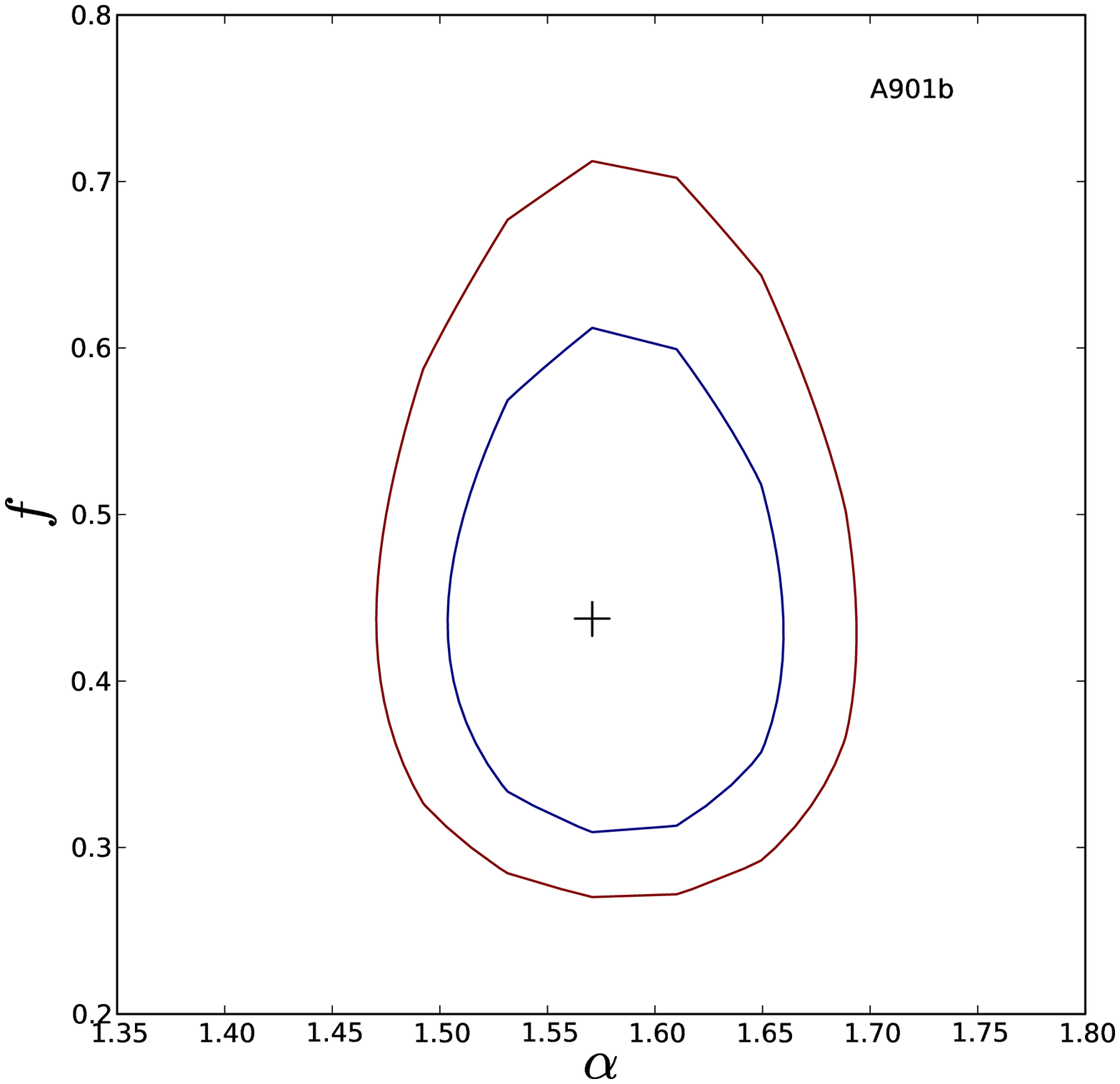}\includegraphics[scale=0.3]{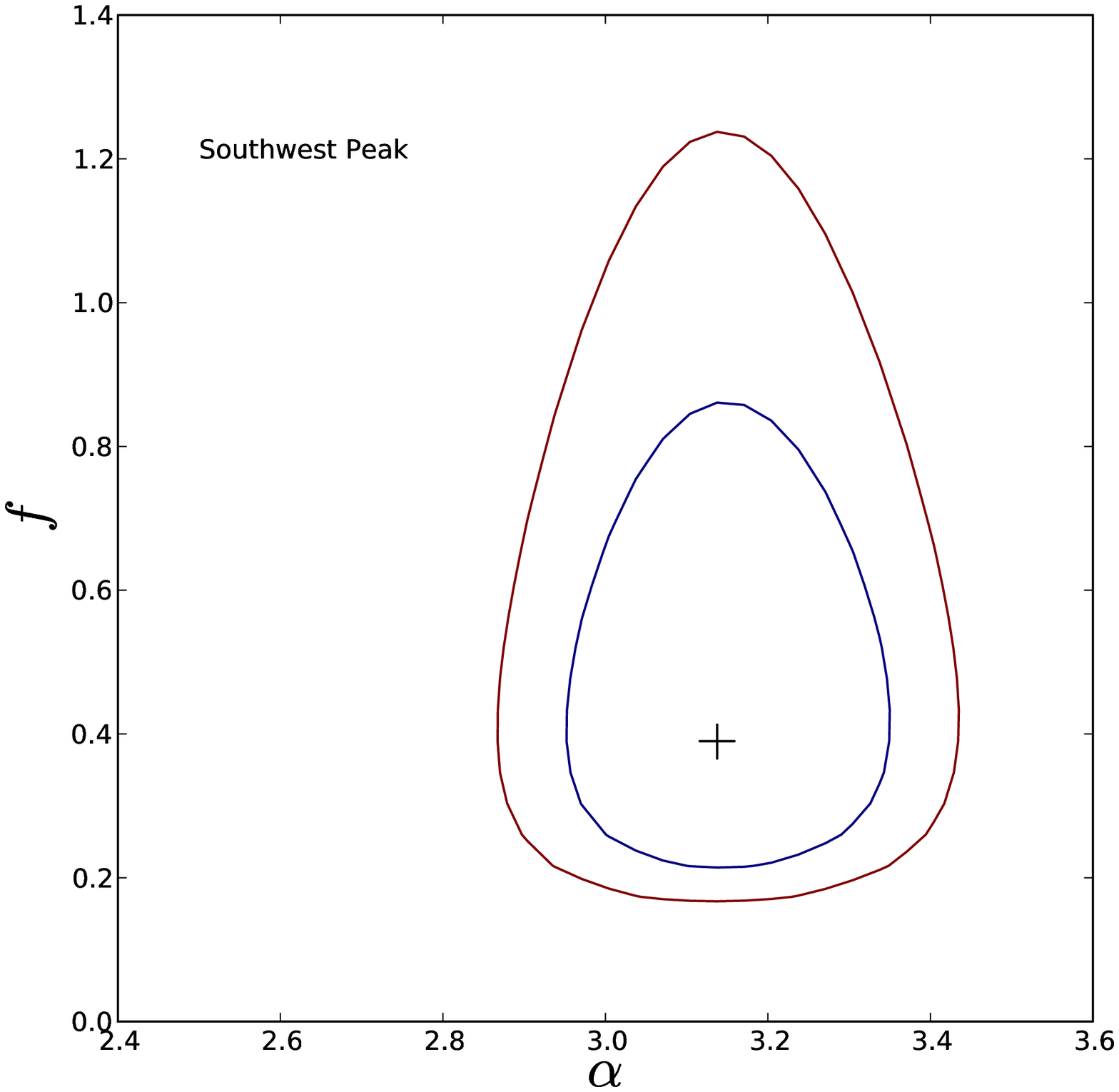}
\caption{ The two panels are the error ellipses  for A901b and the Southwest Peak derived from the parametric modeling, described in  \S~\ref{sec:pm} . The y-axis $f$ is the axis ratio and the x-axis $\alpha$ is the position angle defined in section \S~\ref{sec:pm} in radians. For both plots we have plotted the $1\sigma$ and $2\sigma$ contours.  Both peaks have non-zero ellipticity at $2\sigma$ level. }
\label{fig:error_param}
\end{figure}

From the non-parametric reconstruction it is clear that A901a and A902 are double peaked systems, while A901b and the SouthWest Peak are closer to a single halo with substructure. We have fitted this data to a singular isothermal ellipse described in \S~\ref{sec:pm}. We fit a single halo centered on the BCG in each cluster. 
Since the aim of this study is to measure the ellipticity of the lens we do this fit for the SouthWest Peak and A901b only because A901a and A902 have irregular structure and hence cannot be modeled as an ellipse. The field of view contains four distinct peaks we consider a patch of 178 square arcminutes around each peak. This ensures that the shear signal is not contaminated by the other dark matter halos.
The constraints on the velocity dispersion, ellipticity and position angle is listed in Table~\ref{tb:data_ellip}.  The constraints on the velocity dispersion are consistent with \cite{2002ApJ...568..141G} within error bars.
We estimate the ellipticity of A901b to be $0.39\pm0.09$, and a position angle to be $90^{\circ}$ implying that the dark matter halo is elliptical in the vertical direction.  The dark matter map of the southwest peak is clearly not spherical. This peak is well fitted by the singular isothermal ellipse and the ellipticity of this peak is $0.4^{+0.13}_{-0.16}$. In Figure~\ref{fig:error_param} we have plotted the joint $1(2)-\sigma$ error probability distribution between the axis ratio and the position angle.
It is not possible to constrain the ellipticity of A901a and A902 with an elliptical mass model. In fact a parametric fit is consistent with the spherical model with high error bars. It is clear from the mass reconstruction that the peaks are disturbed. As a matter of fact A901a has two distinct peaks, the second peak coincides with an infalling X-ray group. A902 also has another galaxy group in the background at a redshift $z=0.46$.

\begin{figure}[h!]
\centering
\includegraphics[scale=0.5]{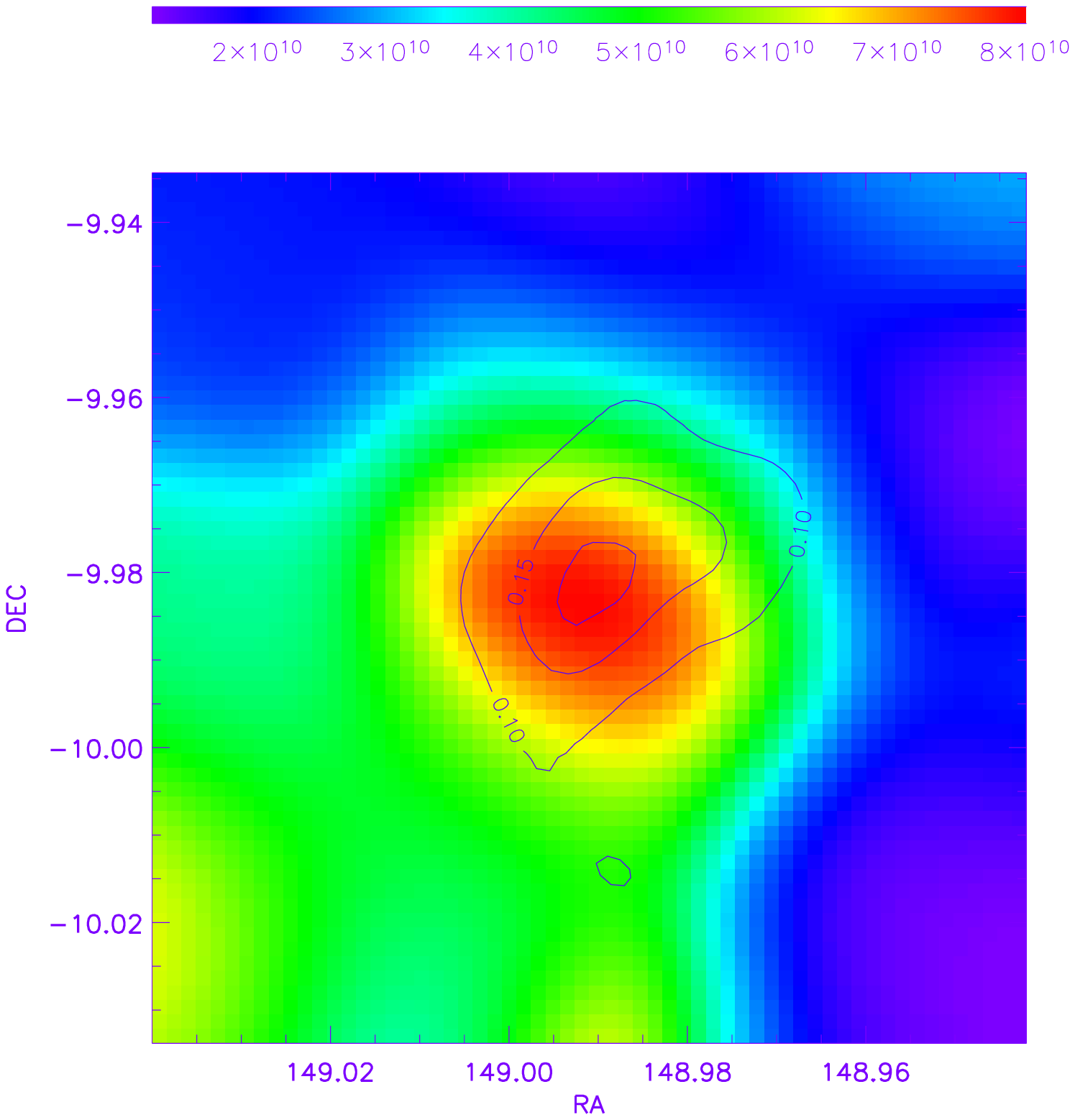}\includegraphics[scale=0.5]{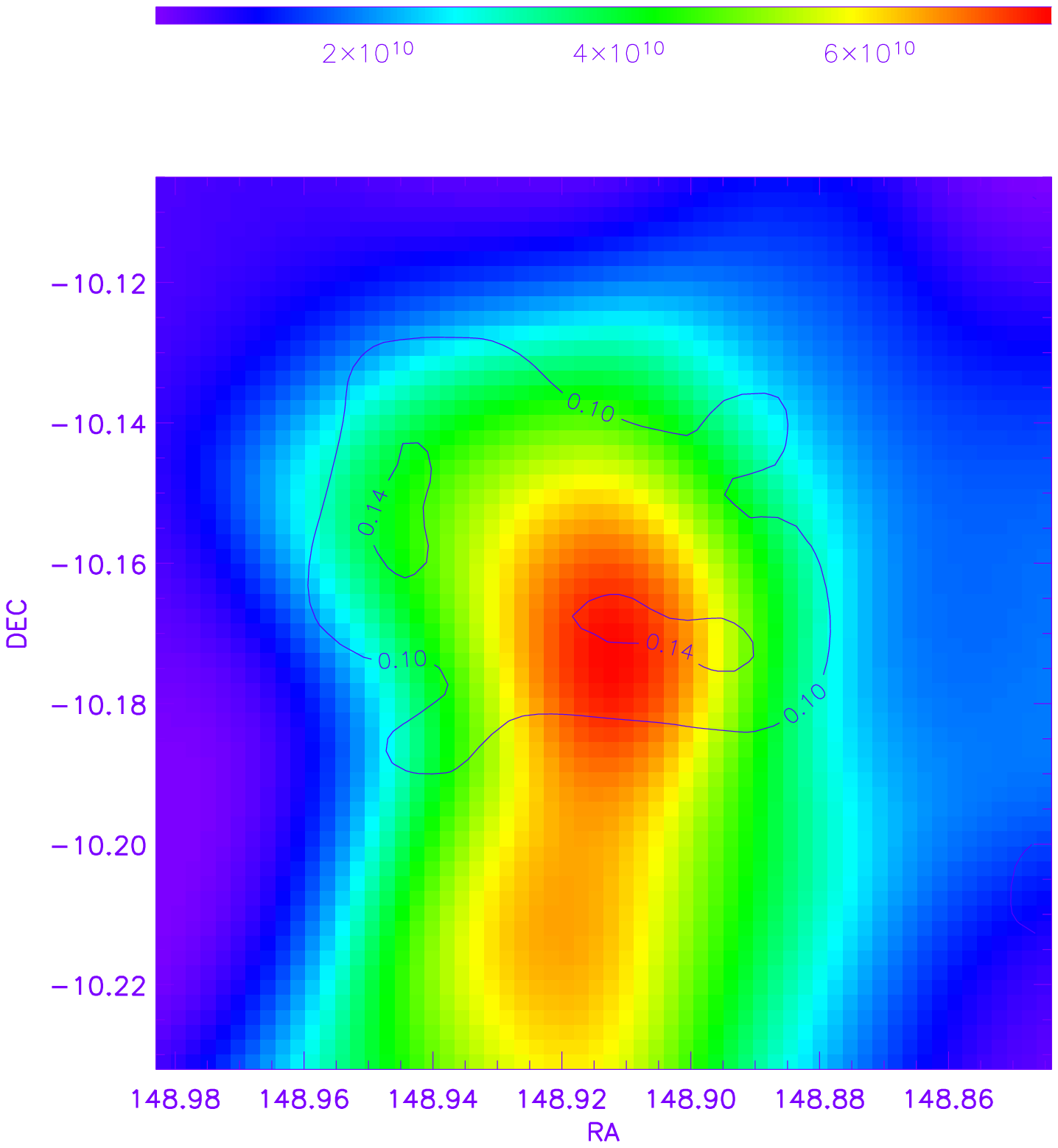}
\caption{ A comparison of the light distribution vs the dark matter distribution. The colors represent the light distribution and the over-layed contours represent the dark matter distribution. Left Panel: A901b. Right Panel: Southwest group. The units of the light distribution is $M_\odot$.}
\label{fig:light}
\end{figure}

\begin{deluxetable}{cccccc}
\tablecolumns{6} \tablewidth{0pc} \tablecaption{Measuring ellipticity of dark matter and light distribution.
\label{tb:data_ellip}}
\tablehead{ \colhead{Peak}  & \colhead{$\sigma_v (km/sec) $}  & \colhead {$ f$} &\colhead{$\alpha $(degrees)}&\colhead{$M<1^{\prime}\; (h_{70}^{-1} 10^{13} M_\odot)$} &\colhead{$\chi_\nu^2$}
}
\startdata
\sidehead{Dark Matter: Parametric }
A901b&$518^{+149}_{-157} $&$0.437^{+0.1}_{-0.087}$&$90.0^{+2.25}_{-2.25}$&$2.66^{+0.29}_{-0.24}$&1.49\\
SW group &$307^{+117}_{-143}$ & $0.42^{+0.18}_{-0.12}$&$180.0^{+7.73}_{-5.15}$ & $0.92^{+0.26}_{-0.22}$&1.46 \\

\cutinhead{Dark Matter: Non-parametric}
A901b&$-$&$0.37^{+0.1}_{-0.1}$&$91.4^{+8.2}_{-8.2}$ &$1.49\pm0.21$&$-$\\
SW group &$-$ & $0.54^{+0.08}_{-0.09}$&$120.0^{+4.8}_{-4.8}$ & $1.31\pm0.12$&$-$ \\

\cutinhead{Light Distribution}
A901b&$-$&$0.81^{+0.1}_{-0.09}$& $130.5^{+12.0}_{-12.0}$&$-$&$-$\\
SW group &$-$ & $0.70^{+0.06}_{-0.05}$&$91.0^{+6.0}_{-6.0}$ & $-$&$-$ \\

\enddata
\end{deluxetable}

\subsection{Comparison between dark matter and light distribution}
We have measured the ellipticity of the light distribution by measuring quadrupole  moments of the cluster member galaxies weighted by their stellar masses. 
The results are listed in Table~\ref{tb:data_ellip}. The light distribution and the dark matter distribution is coincident for A901b. In the Southwest peak the light distribution is coincident with one of the sub-peaks. 
The light distribution is less elliptical compared to the mass distribution. The results from the non-parametric reconstruction of the Southwest group indicates that the major axis of the light distribution and the dark matter distribution are not coincident. This is because the major axis of the Southwest peak is influenced by the smaller sub-peaks whereas the light distrbution has a single peak. Figure~\ref{fig:light} shows a comparison of the light and dark matter distribution. The background map represents the stellar mass of the cluster member in units of $M_\odot$. The contours represent the dark matter distribution.

\subsection{Comparison between parametric and non-parametric results}
We have computed the ellipticity and the position angle for A901b and the Southwest Peak parametrically and non-parametrically. The results for A901b using both methods are consistent. The position angle for the Southwest Peak is inconsistent between the measurements. This peak has an irregular mass distribution, hence an elliptical mass model does not provide a full description of its shape.
The values of the measured ellipticity are very close to that expected from simulations \citep{2006ApJ...647....8H, 2005ApJ...627..647B} and previous observations of dark matter galaxy halo ellipticities from the Red Sequence Cluster Survey(RCS) \cite{2004ApJ...606...67H}  survey. 
The measured velocity dispersion for the peaks are consistent with the results from ground based investigations \citep{2002ApJ...568..141G}. The error in the minor to major axis ratio is quite high since weak shear is a noisy estimator of the dark matter distribution, however it is the only way to uniquely measure the dark matter halo shape.
 
 
\section{Discussion and Future Work}
\label{sec:dis}
In this paper, we have measured the projected ellipticity of dark matter halos non-parametrically.
This is done using an improved PBL by including smoothing of the ellipticity field. We have also calculated the covariance of the resulting mass distribution making the errors of the reconstruction well understood. We have applied this technique to the super cluster A901/902 and reconstructed each of the peaks individually and measured the ellipticity of A901b and Southwest Peak  from the PBL reconstruction and using parametric models. The other two peaks A901a and A902 have a lot of substructure and cannot not be modeled as ellipses. 
We have not considered the line-of-sight ellipticity of the dark matter halos of Abell 901/902. This is because lensing probes the projected mass distribution. \cite{2009MNRAS.393.1235C} have fitted triaxial NFW to galaxy clusters with high errors on the concentration and mass suggesting that more information form strong lensing, X-rays and Sunyaev Zeldovich effect is required to constrain parameters pertaining to the line-of-sight. A joint analysis using X-rays and lensing has been done by \cite{2009arXiv0912.2648M} to constrain the ellipticity along the line of sight.

In future we will measure the ellipticity of the projected dark matter halos for a large sample of clusters and compare them to simulations \citep{2005ApJ...618....1H}. The error introduced due to the line-of-sight ellipticity for a large sample will be insignificant. A similar study has been done 
 in X-rays by  \cite{2005ApJ...624..606J} by studying the ratio of the higher order moments to the zeroth order moments. 

\section{Acknowledgements}
SD would like to thank Kristian Pedersen, Signe Riemer Sorensen, Sarah Bridle, John Paul Kneib, Matthias Bartelmann, Peter Schneider,  Julien Merten, Marceau Limousin, Ardis Eliasdottir  and Steve McMillan for useful suggestions. This work was supported by NSF Award 0908307 and NASA ATP NNG05GF61G. AM is supported by Dark Cosmology Center funded by the Danish National Research Foundation.

\begin{appendix}
\section{Inverting the Covariance matrix for $\chi^2$ minimization.}
\label{sec:svd}
In order to do a $\chi^2$ minimization the inverse of the covariance matrix is important. Before doing the inverse it is important to understand the properties of the covariance matrix. It is essential to remember that every image location is not independent. This is because we are smoothing the data, this makes the number of image  positions larger than the number of 'resolution units'  for a given smoothing scale. This situation has been encountered before \citep{2005MNRAS.362.1363P,2005MNRAS.361..824G} when the data vector was larger than the number of independent information content. This makes input data degenerate, i.e for a smoothing scale considerably higher than the interparticle separation two positions spatially close to each other essentially represnt the same information. 
 This makes the covariance matrix singular. 
The number of independent components of the covariance matrix is inversely proportional to the area under the smoothing kernel.

We invert this singular matrix by using the Singular Value Decomposition (SVD) implementation of Tikhonov regularization.  The traditional SVD matrix inversion is given by,

\be
{M^{-1}=V S U^{T}}
\label{eqn:svd}
\ee
Here U and V are orthogonal matrices, since C is symmetric U=V. S is a diagonal matrix , the diagonal elements are given by $\{1/s\}$. Here s represents the eigenvalues of C.
In case of a singular matrix some of the eigenvalues are zero. Usually s is written in descending order and the first $l$ non-zero eigen values are used for the matrix inversion. For the $n-l$ eigen values $1/s$ is replaced by zero. In real cases it is commonly seen that the eigen values are not zero, rather they are numerically very small, and hence dominated by round-off error. These problems ar termed as ill-conditioned problems. In order to deal with this situation a truncated SVD is used to do the matrix inversion, i.e eigenvalues below a certain threshold are considered to be zero. Choice of this threshold is dependent on the particular problem. If the singular values of a matrix can be distinguished from the non-singular ones in a well-defined fashion then the choice of the threshold becomes simple. This is the case when there is a substantial gap between the largest singular eigenvalue and the smallest non-singular eigenvalue. 
However, in this particular problem where the covariance is due to gaussian smoothing of the data there is no such distinction. As a matter of fact the eigenvalues smoothly asymptote toward zero. Hence the problem does not direct us toward any obvious choice of the threshold value. Since the problem is severly ill-posed we use Tikhonov regularization(ref) to invert the covariance matrix. This is equivalent to

\be
\mathrm{M^{-1}=V f U^{T}}
\ee

where $f_m=\frac{s_m}{s_m^2+\alpha^2}$ are the filter factors. For $s_m>>\alpha, f_m=1/s_m$ and for $s_m<<\alpha, f_m=0$. The presence of the regularization ensures that there is a smooth transition between $f_m=1/s_m$ and $f_m=0$ instead of an abrupt cutoff threshold. The regularization parameter $\alpha$ is given by,
\be
\alpha= C \zeta^2
\ee
As $\alpha$ decreases more and more eigenvalues are included in matrix inversion.
As the smoothing scale is increased the area becomes higher, $\alpha$ becomes higher and the number of modes used in the matrix inversion becomes lower. 
\section{Generalization to the Non-linear regime}
In the previous section we have evaluated the covariance for the reconstructed $\kappa$ in the linear regime. Inverse techniques like PBL are often able to reconstruct the semi-strong regime of weak lensing clusters within a couple of iterations. Hence we write down a general formalism in the semi-strong region.

In the second iteration the potential is given by,
\be
\psi^{(1)}=(G)^{T}C^{-1} \hat{ \varepsilon}= L(\varepsilon^{(0)}- \hat{\varepsilon})
\label{eqn:psi1}
\ee

where $\varepsilon^{(0)i}$ is the modelled ellipticity from the first step of the minimization. In order to get the correct solution for the potential both components of the ellipticity will be summed, for the sake of simplicity of notation we have not included it in Equation~\ref{eqn:psi1}. We assume that we are in the semi-strong region and $\kappa<1$, hence the modelled ellipticity id given by,

\be
\langle\varepsilon^{(0)}_m\rangle=\frac{\gamma^{(0)}_m}{(1-\kappa^{(0)}_m)}=\gamma^{(0)}_m(1+\kappa^{(0)}_m)
\label{eqn:expand}
\ee

The reconstructed $\kappa$ in the second iteration is given by,

\be
\kappa^{(1)}=K\psi^{(0)}+K\psi^{(1)}
\ee

The expectation value for the ellipticity is given by,
\be
\langle\varepsilon_m\rangle=\gamma^{(1)}_m(1+\kappa^{(1)}_m)
\ee

The covariance in $\kappa$ after the first iteration is given by,

\be
\cov^{\kappa,(1)}=\langle K\psi^{(0)}\psi^{(0)T} K^T+2 K\psi^{(1)}\psi^{(0)T} K^T+K\psi^{(1)}\psi^{(1)T} K^T\rangle
\ee

We have already calculated the covariance due to the first term in the previous section.  We have to evaluate the second term and the third term. The second term is given by

\be
term2=2 K L( \langle\hat{\varepsilon}\varepsilon^{(0)}\rangle-\langle\hat{\varepsilon} \hat{\varepsilon}\rangle )L^{T} K^{T}
\ee

and the third term is given by,

\be
term3= K L( \langle{\varepsilon^{(0)}}\varepsilon^{(0)}\rangle+\langle\hat{\varepsilon} \hat{\varepsilon}\rangle-2\langle\hat{\varepsilon}\varepsilon^{(0)}\rangle )L^{T} K^{T}
\ee

Now we will evaluate the covariance the terms $\langle{\varepsilon^{(0)}}\varepsilon^{(0)}\rangle, \langle\hat{\varepsilon} \hat{\varepsilon}\rangle, \langle\hat{\varepsilon}\varepsilon^{(0)}\rangle$ .

\begin{eqnarray}
\langle\varepsilon^{(0)}_m\varepsilon^{(0)}_n\rangle=\gamma^{(0)}_m(1+\kappa^{(0)}_m)\gamma_n^{(0)}(1+\kappa^{(0)}_n)\\
\langle\hat{\varepsilon}_m \hat{\varepsilon}_n\rangle=Q_{mp}Q_{nq}\langle\varepsilon_m\varepsilon_n\rangle=Q_{mp}Q_{nq}\gamma^{(1)}_p(1+\kappa^{(1)}_p)\gamma_q^{(1)}(1+\kappa^{(1)}_q)\\
 \langle\hat{\varepsilon}_m\varepsilon^{(0)}_n\rangle=Q_{mp}\gamma^{(1)}_p(1+\kappa^{(1)}_p)\gamma_n^{(0)}(1+\kappa^{(0)}_n)
\end{eqnarray}

Using these expressions for term2 and term3 we can evaluate the covariance in the reconstructed mass after two iterations. We can do a similar calculation for the next steps iteratively and keep higher order terms in the expansion of Equation~\ref{eqn:expand} and obtain an expression for the covariance of the reconstructed mass.

\end{appendix}

\begin{thebibliography}{83}
\expandafter\ifx\csname natexlab\endcsname\relax\def\natexlab#1{#1}\fi

\bibitem[{{Allen}(1998)}]{1998MNRAS.296..392A}
{Allen}, S.~W. 1998, \mnras, 296, 392

\bibitem[{{Bacon} {et~al.}(2006){Bacon}, {Goldberg}, {Rowe}, \&
  {Taylor}}]{2006MNRAS.365..414B}
{Bacon}, D.~J., {Goldberg}, D.~M., {Rowe}, B.~T.~P., \& {Taylor}, A.~N. 2006,
  \mnras, 365, 414

\bibitem[{{Bailin} \& {Steinmetz}(2005)}]{2005ApJ...627..647B}
{Bailin}, J., \& {Steinmetz}, M. 2005, \apj, 627, 647

\bibitem[{{Bardeau} {et~al.}(2007){Bardeau}, {Soucail}, {Kneib}, {Czoske},
  {Ebeling}, {Hudelot}, {Smail}, \& {Smith}}]{2007A&A...470..449B}
{Bardeau}, S., {Soucail}, G., {Kneib}, J.-P., {Czoske}, O., {Ebeling}, H.,
  {Hudelot}, P., {Smail}, I., \& {Smith}, G.~P. 2007, \aap, 470, 449

\bibitem[{{Bartelmann}(1995)}]{1995A&A...303..643B}
{Bartelmann}, M. 1995, \aap, 303, 643

\bibitem[{{Bartelmann} {et~al.}(2006){Bartelmann}, {Doran}, \&
  {Wetterich}}]{2006A&A...454...27B}
{Bartelmann}, M., {Doran}, M., \& {Wetterich}, C. 2006, \aap, 454, 27

\bibitem[{{Brada{\v c}} {et~al.}(2006){Brada{\v c}}, {Clowe}, {Gonzalez},
  {Marshall}, {Forman}, {Jones}, {Markevitch}, {Randall}, {Schrabback}, \&
  {Zaritsky}}]{2006ApJ...652..937B}
{Brada{\v c}}, M., {Clowe}, D., {Gonzalez}, A.~H., {Marshall}, P., {Forman},
  W., {Jones}, C., {Markevitch}, M., {Randall}, S., {Schrabback}, T., \&
  {Zaritsky}, D. 2006, \apj, 652, 937

\bibitem[{{Brada{\v c}} {et~al.}(2005){Brada{\v c}}, {Schneider}, {Lombardi},
  \& {Erben}}]{2005A&A...437...39B}
{Brada{\v c}}, M., {Schneider}, P., {Lombardi}, M., \& {Erben}, T. 2005, \aap,
  437, 39

\bibitem[{{Brada{\v c}} {et~al.}(2009){Brada{\v c}}, {Treu}, {Applegate},
  {Gonzalez}, {Clowe}, {Forman}, {Jones}, {Marshall}, {Schneider}, \&
  {Zaritsky}}]{2009ApJ...706.1201B}
{Brada{\v c}}, M., {Treu}, T., {Applegate}, D., {Gonzalez}, A.~H., {Clowe}, D.,
  {Forman}, W., {Jones}, C., {Marshall}, P., {Schneider}, P., \& {Zaritsky}, D.
  2009, \apj, 706, 1201

\bibitem[{{Bridle} {et~al.}(1998){Bridle}, {Hobson}, {Lasenby}, \&
  {Saunders}}]{1998MNRAS.299..895B}
{Bridle}, S.~L., {Hobson}, M.~P., {Lasenby}, A.~N., \& {Saunders}, R. 1998,
  \mnras, 299, 895

\bibitem[{{Broadhurst} {et~al.}(2005){Broadhurst}, {Takada}, {Umetsu}, {Kong},
  {Arimoto}, {Chiba}, \& {Futamase}}]{2005ApJ...619L.143B}
{Broadhurst}, T., {Takada}, M., {Umetsu}, K., {Kong}, X., {Arimoto}, N.,
  {Chiba}, M., \& {Futamase}, T. 2005, \apjl, 619, L143

\bibitem[{{Broadhurst} {et~al.}(2008){Broadhurst}, {Umetsu}, {Medezinski},
  {Oguri}, \& {Rephaeli}}]{2008ApJ...685L...9B}
{Broadhurst}, T., {Umetsu}, K., {Medezinski}, E., {Oguri}, M., \& {Rephaeli},
  Y. 2008, \apjl, 685, L9

\bibitem[{{Buote} \& {Tsai}(1995)}]{1995ApJ...452..522B}
{Buote}, D.~A., \& {Tsai}, J.~C. 1995, \apj, 452, 522

\bibitem[{{Clowe} {et~al.}(2006){Clowe}, {Brada{\v c}}, {Gonzalez},
  {Markevitch}, {Randall}, {Jones}, \& {Zaritsky}}]{2006ApJ...648L.109C}
{Clowe}, D., {Brada{\v c}}, M., {Gonzalez}, A.~H., {Markevitch}, M., {Randall},
  S.~W., {Jones}, C., \& {Zaritsky}, D. 2006, \apjl, 648, L109

\bibitem[{{Clowe} \& {Schneider}(2001)}]{2001A&A...379..384C}
{Clowe}, D., \& {Schneider}, P. 2001, \aap, 379, 384

\bibitem[{{Corless} {et~al.}(2009){Corless}, {King}, \&
  {Clowe}}]{2009MNRAS.393.1235C}
{Corless}, V.~L., {King}, L.~J., \& {Clowe}, D. 2009, \mnras, 393, 1235

\bibitem[{{Cypriano} {et~al.}(2001){Cypriano}, {Sodr{\'e}}, {Campusano},
  {Kneib}, {Giovanelli}, {Haynes}, {Dale}, \& {Hardy}}]{2001AJ....121...10C}
{Cypriano}, E.~S., {Sodr{\'e}}, L.~J., {Campusano}, L.~E., {Kneib}, J.-P.,
  {Giovanelli}, R., {Haynes}, M.~P., {Dale}, D.~A., \& {Hardy}, E. 2001, \aj,
  121, 10

\bibitem[{{Cypriano} {et~al.}(2004){Cypriano}, {Sodr{\'e}}, {Kneib}, \&
  {Campusano}}]{2004ApJ...613...95C}
{Cypriano}, E.~S., {Sodr{\'e}}, L.~J., {Kneib}, J., \& {Campusano}, L.~E. 2004,
  \apj, 613, 95

\bibitem[{{Deb} {et~al.}(2008){Deb}, {Goldberg}, \&
  {Ramdass}}]{2008ApJ...687...39D}
{Deb}, S., {Goldberg}, D.~M., \& {Ramdass}, V.~J. 2008, \apj, 687, 39

\bibitem[{{Diego} {et~al.}(2007){Diego}, {Tegmark}, {Protopapas}, \&
  {Sandvik}}]{2007MNRAS.375..958D}
{Diego}, J.~M., {Tegmark}, M., {Protopapas}, P., \& {Sandvik}, H.~B. 2007,
  \mnras, 375, 958

\bibitem[{{Eifler} {et~al.}(2008{\natexlab{a}}){Eifler}, {Kilbinger}, \&
  {Schneider}}]{2008A&A...482....9E}
{Eifler}, T., {Kilbinger}, M., \& {Schneider}, P. 2008{\natexlab{a}}, \aap,
  482, 9

\bibitem[{{Eifler} {et~al.}(2008{\natexlab{b}}){Eifler}, {Schneider}, \&
  {Hartlap}}]{2008arXiv0810.4254E}
{Eifler}, T., {Schneider}, P., \& {Hartlap}, J. 2008{\natexlab{b}}, ArXiv
  e-prints

\bibitem[{{Eke} {et~al.}(1996){Eke}, {Cole}, \& {Frenk}}]{1996MNRAS.282..263E}
{Eke}, V.~R., {Cole}, S., \& {Frenk}, C.~S. 1996, \mnras, 282, 263

\bibitem[{{Francis} {et~al.}(2009){Francis}, {Lewis}, \&
  {Linder}}]{2009MNRAS.393L..31F}
{Francis}, M.~J., {Lewis}, G.~F., \& {Linder}, E.~V. 2009, \mnras, 393, L31

\bibitem[{{Frenk} {et~al.}(1990){Frenk}, {White}, {Efstathiou}, \&
  {Davis}}]{1990ApJ...351...10F}
{Frenk}, C.~S., {White}, S.~D.~M., {Efstathiou}, G., \& {Davis}, M. 1990, \apj,
  351, 10

\bibitem[{{Gazta{\~n}aga} \& {Scoccimarro}(2005)}]{2005MNRAS.361..824G}
{Gazta{\~n}aga}, E., \& {Scoccimarro}, R. 2005, \mnras, 361, 824

\bibitem[{{Geiger} \& {Schneider}(1999)}]{1999MNRAS.302..118G}
{Geiger}, B., \& {Schneider}, P. 1999, \mnras, 302, 118

\bibitem[{{Goldberg} {et~al.}(2009){Goldberg}, {Chessey}, {Harris}, \&
  {Richards}}]{2009arXiv0912.0916G}
{Goldberg}, D.~M., {Chessey}, M.~K., {Harris}, W.~B., \& {Richards}, G.~T.
  2009, ArXiv e-prints

\bibitem[{{Gray} {et~al.}(2002){Gray}, {Taylor}, {Meisenheimer}, {Dye}, {Wolf},
  \& {Thommes}}]{2002ApJ...568..141G}
{Gray}, M.~E., {Taylor}, A.~N., {Meisenheimer}, K., {Dye}, S., {Wolf}, C., \&
  {Thommes}, E. 2002, \apj, 568, 141

\bibitem[{{Gray} {et~al.}(2009){Gray}, {Wolf}, {Barden}, {Peng},
  {H{\"a}u{\ss}ler}, {Bell}, {McIntosh}, {Guo}, {Caldwell}, {Bacon}, {Balogh},
  {Barazza}, {B{\"o}hm}, {Heymans}, {Jahnke}, {Jogee}, {van Kampen}, {Lane},
  {Meisenheimer}, {S{\'a}nchez}, {Taylor}, {Wisotzki}, {Zheng}, {Green},
  {Beswick}, {Saikia}, {Gilmour}, {Johnson}, \&
  {Papovich}}]{2009MNRAS.393.1275G}
{Gray}, M.~E., {Wolf}, C., {Barden}, M., {Peng}, C.~Y., {H{\"a}u{\ss}ler}, B.,
  {Bell}, E.~F., {McIntosh}, D.~H., {Guo}, Y., {Caldwell}, J.~A.~R., {Bacon},
  D., {Balogh}, M., {Barazza}, F.~D., {B{\"o}hm}, A., {Heymans}, C., {Jahnke},
  K., {Jogee}, S., {van Kampen}, E., {Lane}, K., {Meisenheimer}, K.,
  {S{\'a}nchez}, S.~F., {Taylor}, A., {Wisotzki}, L., {Zheng}, X., {Green},
  D.~A., {Beswick}, R.~J., {Saikia}, D.~J., {Gilmour}, R., {Johnson}, B.~D., \&
  {Papovich}, C. 2009, \mnras, 393, 1275

\bibitem[{{Heymans} {et~al.}(2006){Heymans}, {Bell}, {Rix}, {Barden}, {Borch},
  {Caldwell}, {McIntosh}, {Meisenheimer}, {Peng}, {Wolf}, {Beckwith},
  {H{\"a}u{\ss}ler}, {Jahnke}, {Jogee}, {S{\'a}nchez}, {Somerville}, \&
  {Wisotzki}}]{2006MNRAS.371L..60H}
{Heymans}, C., {Bell}, E.~F., {Rix}, H.-W., {Barden}, M., {Borch}, A.,
  {Caldwell}, J.~A.~R., {McIntosh}, D.~H., {Meisenheimer}, K., {Peng}, C.~Y.,
  {Wolf}, C., {Beckwith}, S.~V.~W., {H{\"a}u{\ss}ler}, B., {Jahnke}, K.,
  {Jogee}, S., {S{\'a}nchez}, S.~F., {Somerville}, R., \& {Wisotzki}, L. 2006,
  \mnras, 371, L60

\bibitem[{{Heymans} {et~al.}(2005){Heymans}, {Brown}, {Barden}, {Caldwell},
  {Jahnke}, {Peng}, {Rix}, {Taylor}, {Beckwith}, {Bell}, {Borch},
  {H{\"a}u{\ss}ler}, {Jogee}, {McIntosh}, {Meisenheimer}, {S{\'a}nchez},
  {Somerville}, {Wisotzki}, \& {Wolf}}]{2005MNRAS.361..160H}
{Heymans}, C., {Brown}, M.~L., {Barden}, M., {Caldwell}, J.~A.~R., {Jahnke},
  K., {Peng}, C.~Y., {Rix}, H.-W., {Taylor}, A., {Beckwith}, S.~V.~W., {Bell},
  E.~F., {Borch}, A., {H{\"a}u{\ss}ler}, B., {Jogee}, S., {McIntosh}, D.~H.,
  {Meisenheimer}, K., {S{\'a}nchez}, S.~F., {Somerville}, R., {Wisotzki}, L.,
  \& {Wolf}, C. 2005, \mnras, 361, 160

\bibitem[{{Heymans} {et~al.}(2008){Heymans}, {Gray}, {Peng}, {van Waerbeke},
  {Bell}, {Wolf}, {Bacon}, {Balogh}, {Barazza}, {Barden}, {B{\"o}hm},
  {Caldwell}, {H{\"a}u{\ss}ler}, {Jahnke}, {Jogee}, {van Kampen}, {Lane},
  {McIntosh}, {Meisenheimer}, {Mellier}, {S{\'a}nchez}, {Taylor}, {Wisotzki},
  \& {Zheng}}]{2008MNRAS.385.1431H}
{Heymans}, C., {Gray}, M.~E., {Peng}, C.~Y., {van Waerbeke}, L., {Bell}, E.~F.,
  {Wolf}, C., {Bacon}, D., {Balogh}, M., {Barazza}, F.~D., {Barden}, M.,
  {B{\"o}hm}, A., {Caldwell}, J.~A.~R., {H{\"a}u{\ss}ler}, B., {Jahnke}, K.,
  {Jogee}, S., {van Kampen}, E., {Lane}, K., {McIntosh}, D.~H., {Meisenheimer},
  K., {Mellier}, Y., {S{\'a}nchez}, S.~F., {Taylor}, A.~N., {Wisotzki}, L., \&
  {Zheng}, X. 2008, \mnras, 385, 1431

\bibitem[{{Ho} {et~al.}(2006){Ho}, {Bahcall}, \& {Bode}}]{2006ApJ...647....8H}
{Ho}, S., {Bahcall}, N., \& {Bode}, P. 2006, \apj, 647, 8

\bibitem[{{Hoekstra} {et~al.}(2004){Hoekstra}, {Yee}, \&
  {Gladders}}]{2004ApJ...606...67H}
{Hoekstra}, H., {Yee}, H.~K.~C., \& {Gladders}, M.~D. 2004, \apj, 606, 67

\bibitem[{{Hopkins} {et~al.}(2005){Hopkins}, {Bahcall}, \&
  {Bode}}]{2005ApJ...618....1H}
{Hopkins}, P.~F., {Bahcall}, N.~A., \& {Bode}, P. 2005, \apj, 618, 1

\bibitem[{{Irgens} {et~al.}(2002){Irgens}, {Lilje}, {Dahle}, \&
  {Maddox}}]{2002ApJ...579..227I}
{Irgens}, R.~J., {Lilje}, P.~B., {Dahle}, H., \& {Maddox}, S.~J. 2002, \apj,
  579, 227

\bibitem[{{Jeltema} {et~al.}(2005){Jeltema}, {Canizares}, {Bautz}, \&
  {Buote}}]{2005ApJ...624..606J}
{Jeltema}, T.~E., {Canizares}, C.~R., {Bautz}, M.~W., \& {Buote}, D.~A. 2005,
  \apj, 624, 606

\bibitem[{{Jing} \& {Suto}(2002)}]{2002ApJ...574..538J}
{Jing}, Y.~P., \& {Suto}, Y. 2002, \apj, 574, 538

\bibitem[{{Kaiser}(1995)}]{1995ApJ...439L...1K}
{Kaiser}, N. 1995, \apjl, 439, L1

\bibitem[{{Kaiser} \& {Squires}(1993)}]{1993ApJ...404..441K}
{Kaiser}, N., \& {Squires}, G. 1993, \apj, 404, 441

\bibitem[{{Kaiser} {et~al.}(1995){Kaiser}, {Squires}, \&
  {Broadhurst}}]{1995ApJ...449..460K}
{Kaiser}, N., {Squires}, G., \& {Broadhurst}, T. 1995, \apj, 449, 460

\bibitem[{{King} \& {Corless}(2007)}]{2007MNRAS.374L..37K}
{King}, L., \& {Corless}, V. 2007, \mnras, 374, L37

\bibitem[{{King} \& {Schneider}(2001)}]{2001A&A...369....1K}
{King}, L.~J., \& {Schneider}, P. 2001, \aap, 369, 1

\bibitem[{{Komatsu} {et~al.}(2009){Komatsu}, {Dunkley}, {Nolta}, {Bennett},
  {Gold}, {Hinshaw}, {Jarosik}, {Larson}, {Limon}, {Page}, {Spergel},
  {Halpern}, {Hill}, {Kogut}, {Meyer}, {Tucker}, {Weiland}, {Wollack}, \&
  {Wright}}]{2009ApJS..180..330K}
{Komatsu}, E., {Dunkley}, J., {Nolta}, M.~R., {Bennett}, C.~L., {Gold}, B.,
  {Hinshaw}, G., {Jarosik}, N., {Larson}, D., {Limon}, M., {Page}, L.,
  {Spergel}, D.~N., {Halpern}, M., {Hill}, R.~S., {Kogut}, A., {Meyer}, S.~S.,
  {Tucker}, G.~S., {Weiland}, J.~L., {Wollack}, E., \& {Wright}, E.~L. 2009,
  \apjs, 180, 330

\bibitem[{{Kormann} {et~al.}(1994){Kormann}, {Schneider}, \&
  {Bartelmann}}]{1994A&A...284..285K}
{Kormann}, R., {Schneider}, P., \& {Bartelmann}, M. 1994, \aap, 284, 285

\bibitem[{{Limousin} {et~al.}(2008){Limousin}, {Richard}, {Kneib}, {Brink},
  {Pell{\'o}}, {Jullo}, {Tu}, {Sommer-Larsen}, {Egami}, {Micha{\l}owski},
  {Cabanac}, \& {Stark}}]{2008A&A...489...23L}
{Limousin}, M., {Richard}, J., {Kneib}, J.-P., {Brink}, H., {Pell{\'o}}, R.,
  {Jullo}, E., {Tu}, H., {Sommer-Larsen}, J., {Egami}, E., {Micha{\l}owski},
  M.~J., {Cabanac}, R., \& {Stark}, D.~P. 2008, \aap, 489, 23

\bibitem[{{Lombardi} \& {Bertin}(1999)}]{1999A&A...348...38L}
{Lombardi}, M., \& {Bertin}, G. 1999, \aap, 348, 38

\bibitem[{{Lombardi} \& {Schneider}(2001)}]{2001A&A...373..359L}
{Lombardi}, M., \& {Schneider}, P. 2001, \aap, 373, 359

\bibitem[{{Lombardi} \& {Schneider}(2002)}]{2002A&A...392.1153L}
---. 2002, \aap, 392, 1153

\bibitem[{{Lombardi} \& {Schneider}(2003)}]{2003A&A...407..385L}
---. 2003, \aap, 407, 385

\bibitem[{{Mandelbaum} {et~al.}(2009){Mandelbaum}, {Seljak}, {Baldauf}, \&
  {Smith}}]{2009arXiv0911.4972M}
{Mandelbaum}, R., {Seljak}, U., {Baldauf}, T., \& {Smith}, R.~E. 2009, ArXiv
  e-prints

\bibitem[{{Mandelbaum} {et~al.}(2008){Mandelbaum}, {Seljak}, \&
  {Hirata}}]{2008JCAP...08..006M}
{Mandelbaum}, R., {Seljak}, U., \& {Hirata}, C.~M. 2008, Journal of Cosmology
  and Astro-Particle Physics, 8, 6

\bibitem[{{Marshall} {et~al.}(2002){Marshall}, {Hobson}, {Gull}, \&
  {Bridle}}]{2002MNRAS.335.1037M}
{Marshall}, P.~J., {Hobson}, M.~P., {Gull}, S.~F., \& {Bridle}, S.~L. 2002,
  \mnras, 335, 1037

\bibitem[{{Mellier}(1999)}]{1999ARA&A..37..127M}
{Mellier}, Y. 1999, \araa, 37, 127

\bibitem[{{Merten} {et~al.}(2009){Merten}, {Cacciato}, {Meneghetti}, {Mignone},
  \& {Bartelmann}}]{2009A&A...500..681M}
{Merten}, J., {Cacciato}, M., {Meneghetti}, M., {Mignone}, C., \& {Bartelmann},
  M. 2009, \aap, 500, 681

\bibitem[{{Morandi} {et~al.}(2009){Morandi}, {Pedersen}, \&
  {Limousin}}]{2009arXiv0912.2648M}
{Morandi}, A., {Pedersen}, K., \& {Limousin}, M. 2009, ArXiv e-prints

\bibitem[{{Oguri} {et~al.}(2009){Oguri}, {Hennawi}, {Gladders}, {Dahle},
  {Natarajan}, {Dalal}, {Koester}, {Sharon}, \&
  {Bayliss}}]{2009arXiv0901.4372O}
{Oguri}, M., {Hennawi}, J.~F., {Gladders}, M.~D., {Dahle}, H., {Natarajan}, P.,
  {Dalal}, N., {Koester}, B.~P., {Sharon}, K., \& {Bayliss}, M. 2009, ArXiv
  e-prints

\bibitem[{{Oguri} {et~al.}(2003){Oguri}, {Lee}, \&
  {Suto}}]{2003ApJ...599....7O}
{Oguri}, M., {Lee}, J., \& {Suto}, Y. 2003, \apj, 599, 7

\bibitem[{{Oguri} {et~al.}(2005){Oguri}, {Takada}, {Umetsu}, \&
  {Broadhurst}}]{2005ApJ...632..841O}
{Oguri}, M., {Takada}, M., {Umetsu}, K., \& {Broadhurst}, T. 2005, \apj, 632,
  841

\bibitem[{{Okabe} {et~al.}(2009){Okabe}, {Takada}, {Umetsu}, {Futamase}, \&
  {Smith}}]{2009arXiv0903.1103O}
{Okabe}, N., {Takada}, M., {Umetsu}, K., {Futamase}, T., \& {Smith}, G.~P.
  2009, ArXiv e-prints

\bibitem[{{Okabe} \& {Umetsu}(2008)}]{2008PASJ...60..345O}
{Okabe}, N., \& {Umetsu}, K. 2008, \pasj, 60, 345

\bibitem[{{Pan} \& {Szapudi}(2005)}]{2005MNRAS.362.1363P}
{Pan}, J., \& {Szapudi}, I. 2005, \mnras, 362, 1363

\bibitem[{{Pedersen} \& {Dahle}(2007)}]{2007ApJ...667...26P}
{Pedersen}, K., \& {Dahle}, H. 2007, \apj, 667, 26

\bibitem[{{Press} \& {Schechter}(1974)}]{1974ApJ...187..425P}
{Press}, W.~H., \& {Schechter}, P. 1974, \apj, 187, 425

\bibitem[{{Rahman} {et~al.}(2006){Rahman}, {Krywult}, {Motl}, {Flin}, \&
  {Shandarin}}]{2006MNRAS.367..838R}
{Rahman}, N., {Krywult}, J., {Motl}, P.~M., {Flin}, P., \& {Shandarin}, S.~F.
  2006, \mnras, 367, 838

\bibitem[{{Rhodes} {et~al.}(2007){Rhodes}, {Massey}, {Albert}, {Collins},
  {Ellis}, {Heymans}, {Gardner}, {Kneib}, {Koekemoer}, {Leauthaud}, {Mellier},
  {Refregier}, {Taylor}, \& {Van Waerbeke}}]{2007ApJS..172..203R}
{Rhodes}, J.~D., {Massey}, R.~J., {Albert}, J., {Collins}, N., {Ellis}, R.~S.,
  {Heymans}, C., {Gardner}, J.~P., {Kneib}, J., {Koekemoer}, A., {Leauthaud},
  A., {Mellier}, Y., {Refregier}, A., {Taylor}, J.~E., \& {Van Waerbeke}, L.
  2007, \apjs, 172, 203

\bibitem[{{Schneider} \& {Bartelmann}(1997)}]{1997MNRAS.286..696S}
{Schneider}, P., \& {Bartelmann}, M. 1997, \mnras, 286, 696

\bibitem[{{Seitz} \& {Schneider}(1995)}]{1995A&A...297..287S}
{Seitz}, C., \& {Schneider}, P. 1995, \aap, 297, 287

\bibitem[{{Seitz} \& {Schneider}(1996)}]{1996A&A...305..383S}
{Seitz}, S., \& {Schneider}, P. 1996, \aap, 305, 383

\bibitem[{{Seitz} \& {Schneider}(1998)}]{1998astro.ph..2051S}
---. 1998, ArXiv Astrophysics e-prints

\bibitem[{{Seitz} \& {Schneider}(2001)}]{2001A&A...374..740S}
---. 2001, \aap, 374, 740

\bibitem[{{Seitz} {et~al.}(1998){Seitz}, {Schneider}, \&
  {Bartelmann}}]{1998A&A...337..325S}
{Seitz}, S., {Schneider}, P., \& {Bartelmann}, M. 1998, \aap, 337, 325

\bibitem[{{Sheth} \& {Tormen}(1999)}]{1999MNRAS.308..119S}
{Sheth}, R.~K., \& {Tormen}, G. 1999, \mnras, 308, 119

\bibitem[{{Spergel} {et~al.}(2007){Spergel}, {Bean}, {Dor{\'e}}, {Nolta},
  {Bennett}, {Dunkley}, {Hinshaw}, {Jarosik}, {Komatsu}, {Page}, {Peiris},
  {Verde}, {Halpern}, {Hill}, {Kogut}, {Limon}, {Meyer}, {Odegard}, {Tucker},
  {Weiland}, {Wollack}, \& {Wright}}]{2007ApJS..170..377S}
{Spergel}, D.~N., {Bean}, R., {Dor{\'e}}, O., {Nolta}, M.~R., {Bennett}, C.~L.,
  {Dunkley}, J., {Hinshaw}, G., {Jarosik}, N., {Komatsu}, E., {Page}, L.,
  {Peiris}, H.~V., {Verde}, L., {Halpern}, M., {Hill}, R.~S., {Kogut}, A.,
  {Limon}, M., {Meyer}, S.~S., {Odegard}, N., {Tucker}, G.~S., {Weiland},
  J.~L., {Wollack}, E., \& {Wright}, E.~L. 2007, \apjs, 170, 377

\bibitem[{{Tyson} {et~al.}(1990){Tyson}, {Wenk}, \&
  {Valdes}}]{1990ApJ...349L...1T}
{Tyson}, J.~A., {Wenk}, R.~A., \& {Valdes}, F. 1990, \apjl, 349, L1

\bibitem[{{Umetsu} \& {Broadhurst}(2008)}]{2008ApJ...684..177U}
{Umetsu}, K., \& {Broadhurst}, T. 2008, \apj, 684, 177

\bibitem[{{van Waerbeke}(2000)}]{2000MNRAS.313..524V}
{van Waerbeke}, L. 2000, \mnras, 313, 524

\bibitem[{{Van Waerbeke} {et~al.}(2000){Van Waerbeke}, {Mellier}, {Erben},
  {Cuillandre}, {Bernardeau}, {Maoli}, {Bertin}, {Mc Cracken}, {Le F{\`e}vre},
  {Fort}, {Dantel-Fort}, {Jain}, \& {Schneider}}]{2000A&A...358...30V}
{Van Waerbeke}, L., {Mellier}, Y., {Erben}, T., {Cuillandre}, J.~C.,
  {Bernardeau}, F., {Maoli}, R., {Bertin}, E., {Mc Cracken}, H.~J., {Le
  F{\`e}vre}, O., {Fort}, B., {Dantel-Fort}, M., {Jain}, B., \& {Schneider}, P.
  2000, \aap, 358, 30

\bibitem[{{Williams} {et~al.}(1999){Williams}, {Navarro}, \&
  {Bartelmann}}]{1999ApJ...527..535W}
{Williams}, L.~L.~R., {Navarro}, J.~F., \& {Bartelmann}, M. 1999, \apj, 527,
  535

\bibitem[{{Wu} \& {Hammer}(1995)}]{1995A&A...299..353W}
{Wu}, X.-P., \& {Hammer}, F. 1995, \aap, 299, 353

\bibitem[{{Zitrin} \& {Broadhurst}(2009)}]{2009arXiv0906.5079Z}
{Zitrin}, A., \& {Broadhurst}, T. 2009, ArXiv e-prints

\bibitem[{{Zitrin} {et~al.}(2009){Zitrin}, {Broadhurst}, {Rephaeli}, \&
  {Sadeh}}]{2009arXiv0907.4232Z}
{Zitrin}, A., {Broadhurst}, T., {Rephaeli}, Y., \& {Sadeh}, S. 2009, ArXiv
  e-prints

\end{thebibliography}

\end{document}